%% file: main.tex
\documentstyle[psfig,longtable,epsf]{mn}

\begin{document}

\input{abstract}

\input{intro}
\input{obs}
\input{results}
\input{gas}
\input{conc}
\input{bib}


\end{document}

%% file: abstract.tex














\title[The SCUBA Local Universe Galaxy Survey I]{The SCUBA Local Universe Galaxy Survey I. First Measurements of the Submillimetre Luminosity and Dust Mass Functions}
\author[Loretta Dunne et al.]
{Loretta Dunne$^{1}$, 
Stephen Eales$^{1}$, 
Michael Edmunds$^1$,
Rob Ivison$^2$,\cr 
Paul Alexander$^3$ 
and David L. Clements$^1$\\
$^1$Department of Physics and Astronomy, University of Wales Cardiff,
PO Box 913, Cardiff, CF2 3YB\\
$^2$ Department of Physics and Astronomy, University College London, Gower
Street, London WC1E 6BT\\
$^3$ Department of Astrophysics, Cavendish Laboratory, Madingley Road,
Cambridge CB3 0HE\\ }

\maketitle


\begin{abstract}



This is the first of a series of papers
presenting results from the SCUBA Local Universe Galaxy Survey
(SLUGS), the first statistical survey of the
submillimetre properties of the local universe. As the initial part
of this survey, we have used the SCUBA camera on the James Clerk Maxwell
Telescope to observe 104 galaxies from the IRAS Bright Galaxy Sample.
We present here the 850$\mu$m flux measurements.

The 60, 100, and 850$\mu$m flux densities are well fitted by
single-temperature dust spectral energy distributions, with the sample
mean and standard deviation for the best-fitting temperature being
$T_{\rm{d}}= 35.6\, \pm\, 4.9\,\rm{K}$ and, for the dust emissivity index
$\beta=1.3\, \pm\, 0.2$. The dust temperature was found to correlate with
60$\mu$m luminosity. The low value of $\beta$ may simply mean that
these galaxies contain a significant amount of dust that is colder
than these temperatures. We have estimated dust masses from the
850$\mu$m fluxes and from the fitted temperature, although if a colder
component at around 20 K is present (assuming a $\beta$ of 2) then
the estimated dust masses are a factor of 1.5--3 too low.

We have made the first direct measurements of the submillimetre
luminosity function (LF) and of the dust mass function. Unlike the
{\em IRAS\/} 60$\mu$m LF, these are well fitted by Schechter
functions. The slope of the 850$\mu$m LF at low luminosities is
steeper than $-2$, implying that the LF must flatten at luminosities
lower than we probe here. We show that extrapolating the 60$\mu$m LF
to 850$\mu$m using a single temperature and $\beta$ does not reproduce
the measured submillimetre LF. A population of `cold' galaxies ($T_{\rm{d}} <
25 \rm{K}$) emitting strongly at submillimetre wavelengths would have
been excluded from the $60 \mu$m selected sample. {\em IF\/} such
galaxies do exist then this
estimate of the 850$\mu$m is biased (it is underestimated). Whether
such a population does exist is unknown at present. 

We have correlated many of the global galaxy properties with the
FIR/submillimetre properties. We find that there is a tendency for
less luminous galaxies to contain hotter dust and to have a greater
star-formation efficiency (cf. Young 1999). The average gas-to-dust
ratio for the sample is $581\pm 43$ (using both the atomic and molecular
hydrogen) which is significantly higher than the Galactic value of
160. We believe this discrepancy is likely to be due to a `cold dust'
component at $T_{\rm{d}}\leq 20\, \rm K$. There is a surprisingly
tight correlation between dust mass and the mass of molecular
hydrogen, estimated from CO measurements, with an intrinsic scatter of
$\simeq$50 per cent.
\end{abstract}

\begin{keywords}
dust, extinction;galaxies --  ISM -- galaxies;luminosity
function, mass function -- galaxies;starburst -- infrared;galaxies
-- submillimetre.
\end{keywords}


%% file: intro.tex


\section{Introduction}

With the advent of the SCUBA bolometer array (Holland et al. 1999) on
the James Clerk Maxwell Telescope\footnote{The JCMT is operated by the
Joint Astronomy Center on behalf of the UK Particle Physics and
Astronomy Research Council, the Netherlands Organization for
scientific Research and the Canadian National Research Council.} it
has, for the first time, become feasible to map the submillimetre
(100$\mu$m$<\lambda<$1 mm) emission of a large ($>$100) sample of
galaxies. While this may seem a paltry feat in comparison to surveys
conducted at other wavelengths, submillimetre astronomy has lagged
behind the rest due to the technical difficulties of building
sensitive enough instruments. Until recently, all that existed were a
handful of submillimetre fluxes of nearby galaxies made with single
element bolometers (Eales, Wynn-Williams \& Duncan 1989; Stark et al.
1989; Clements, Andreani
\& Chase 1993). No information about the distribution of the dust was
available and even the fluxes themselves were subject to considerable
uncertainty, especially when beam corrections had to be made in order
to compare with measurements in the far infra-red (FIR :
10$\mu$m$<\lambda<$100$\mu$m) made with the {\em IRAS\/}
satellite. Quite surprisingly, even before SCUBA, there had been some
successful submillimetre observations of high redshift objects (Dunlop
et al. 1994; Isaak et al. 1994; Ivison 1995) but interpretation of the
measurements, particularly the high implied dust masses, was hampered
by a lack of knowledge of local submillimetre properties, such as
luminosity and dust mass functions (Eales \& Edmunds 1996, 1997
hereafter EE96,97). We have started to use SCUBA to carry out a survey
of a large number of nearby galaxies and hope that this first
statistical submillimetre survey will answer a number of questions.

\subsection{How much dust does a galaxy contain?}

The amount of dust in a galaxy is a measure of the quantity of heavy
elements in the interstellar medium (ISM), since $\sim$ 50 per cent of
the heavy elements are locked up in dust and there is some evidence
that this fraction is constant from galaxy to galaxy (EE96). This is a
different and complementary way of investigating the heavy element
content from simply measuring the metallicity, which is of course, the
mass of heavy elements {\em per unit mass\/} of gas not including that
which is locked up in dust. EE96 have shown that the way in which the
dust mass of a galaxy evolves depends critically on the mode of
evolution:-- whether the galaxy can be treated as a closed box or
whether gas is inflowing onto or outflowing from it. Dust masses have
been estimated from {\em IRAS\/} measurements but the resulting
gas-to-dust ratios are a factor of $\sim$ 5--10 higher than for our
own Galaxy (Devereux \& Young 1990; Sanders et al. 1991) suggesting a
problem with the method. The submillimetre offers a number of
advantages for estimating dust masses. At longer submillimetre
wavelengths ($\lambda>350\mu$m) we are sampling the Rayleigh-Jeans
part of the Planck function, where the flux is least sensitive to
temperature and most sensitive to the mass of the emitting
material. This is a good thing because accurate dust temperatures
($T_{\rm{d}}$) are difficult to estimate. Dust radiates as a `grey
body', which is a Planck function modified by an emissivity term
Q$_{em}\propto
\nu^{\beta}$, where the emissivity index ($\beta$) is believed to lie between 1 and 2
(Hildebrand 1983). To derive the temperature, an assumption must be
made about $\beta$ or vice-versa. If an incorrect temperature is
assumed at say 850$\mu$m, then the dust mass will be wrong by
approximately the fractional error in $T_{\rm{d}}$, but at shorter FIR
wavelengths (100$\mu$m) the error will be much larger as flux goes as
$T_{\rm{d}}^{(4+\beta)}$. Generally, using {\em IRAS\/} measurements
alone to determine the temperature and mass will always over-estimate
the first and so under-estimate the second. The reason for this is
that because of the sensitivity to temperature of emission on the Wien
side of the grey-body curve, the FIR emission from a galaxy is
dominated by warm dust, even when there is relatively little mass in
this component. Thus fits of grey-body curves to FIR fluxes are biased
towards higher temperatures than is warranted by the relative masses
of warm and cold dust. This is, of course, the probable explanation of
the high gas-to-dust ratios observed by Devereux \& Young (1990), and
also explains why it has been so difficult to demonstrate the presence
of cold dust in galaxies. Longer wavelength studies (200$\mu$m --
850$\mu$m) using {\em COBE\/}, {\em ISO\/} and SCUBA (Sodroski et al.
1994; Reach et al. 1995; Alton et al. 1998a,b; Davies et al. 1999;
Frayer et al. 1999; Papadopoulos \& Seaquist 1999) have now confirmed
the existence of cold dust components at 15 $< T_{\rm{d}} <$ 25 K in
nearby spiral galaxies as predicted for grains heated by the general
interstellar radiation field (Cox, Kr\"{u}gel \& Mezger 1986), rather
than in star forming regions.

There is one further difficulty in estimating dust masses
at any wavelength:-- our poor knowledge of the dust mass opacity
coefficient $\kappa_d(\nu)$, which is needed to give absolute values
for mass. This varies with frequency in the same way as the
emissivity, and so current measurements at 120--200$\mu$m (Hildebrand
1983, Draine \& Lee 1984) must be extrapolated to the wavelength of
interest using the assumed value of $\beta$, raising the possibility
of large errors since $\beta$ itself is poorly known. Hughes, Dunlop
\& Rawlings (1997) estimate the uncertainty in this coefficient to be a factor
$\sim$8 at 850$\mu$m. For many issues, including galaxy evolution,
absolute values of mass are not important (EE96) provided the relative
masses are correct.

\subsection{How much optical light is absorbed by dust?}

There is an increasing need to understand the effects of dust on our
optical view of the universe. Optical astronomers have recently shown
that the UV luminosity density in the universe may rise from now to
$z\sim1$ and then fall off at higher redshifts (Lilly et al. 1996;
Madau et al. 1996). There have been claims that this represents the
`star-formation history of the universe' but how much is this affected
by dust? In general, how important are selection effects caused by
dust, and is the high redshift universe significantly attenuated by
foreground dusty objects? (Davies et al. 1997). Mapping large numbers of
galaxies at submillimetre wavelengths will help determine how the dust
is distributed relative to the stars, and how effective it is at
absorbing optical light.

\subsection{How much cosmological evolution is seen in the \protect\newline submillimetre?}

With the deep SCUBA surveys now taking place (Smail, Ivison \& Blain
1997; Hughes et al. 1998; Barger et al. 1998,1999; Blain et al. 1999a;
Eales et al. 1999; Lilly et al. 1999), there is now probably more
information about the submillimetre properties of the distant universe
than the local universe. The galaxies in the deep submillimetre
surveys appear very similar to the ULIRGs found nearby (Smail et
al. 1998; Hughes et al. 1998; Lilly et al. 1999). If the dust in these
high-redshift objects is heated by young stars, as seems to be largely
the case for nearby ULIRGs (Genzel et al. 1998) then, since these
sources make up $>$20 per cent of the total extra-galactic background
emission, this implies that $\geq$10 per cent of all the stars that
have ever formed did so in in this kind of extreme object (Eales et
al. 1999). Studies of the cosmological evolution of this population
suggests that it is very similar to that seen at optical wavelengths
in the Lilly-Madau curve.  What is derived for the cosmological
evolution, however, depends critically on what is assumed about the
submillimetre properties of the local universe (cf. results of Eales
et al. 1999 and Blain et al. 1999b). Currently, rather than working
from a local submillimetre luminosity function, most investigations
have perforce started from a local {\em IRAS\/} 60$\mu$m luminosity
function and extrapolated to submillimetre wavelengths using some
assumptions about the average FIR-submm SED. The lack of a direct
submillimetre luminosity function has been a severe limitation when
interpreting the results of deep surveys. Even before SCUBA,
submillimetre observations of high-redshift radio galaxies and quasars
have found unusually high dust masses. Here again, explanation has
been restricted by our ignorance of the local universe, in this case
the statistics of dust masses in galaxies (EE 96).

\subsection{Is CO a good tracer of molecular hydrogen?}

Carbon monoxide (CO) has long been used as a tracer for molecular gas
(H$_2$). The conversion factor between CO and H$_2$ (known as the `$X$'
factor) is quite uncertain
and may be a function of physical environment such as density,
temperature and metallicity (Maloney 1990; Wilson 1995). Dust is an
alternative tracer of H$_2$, and may help support the case for or
against CO. Any variations of the ratio of CO to dust would provide insight
on the way in which the $X$ factor depends on galaxy properties.

\subsection{The scope of the survey}

The ideal way to carry out the survey would be to do a blank field
survey of large parts of the sky and measure the redshifts of all
objects detected, a method which would be largely free of any
selection effects. Unfortunately, this is impractical at the moment
because the field of view of SCUBA is only $\sim $2 arcmins, making
the time needed for such a survey prohibitively long. Instead, we
decided to use the less ideal but more practical method of observing
galaxies drawn from as many different complete samples selected in as
many different wavebands as possible. This procedure still allows us
to produce unbiased estimates of the submillimetre luminosity function
and of the dust mass function using `accessible volume' techniques
(Avni \& Bahcall 1980). These will not be biased unless there is a
class of galaxies which is not represented in any of the
samples. Provided at least one member of such a class of objects is
present in one of the samples then the estimates will be unbiased
although clearly with large random errors.

In this paper we present the results from a sample selected at
60$\mu$m. We present first estimates of the luminosity and dust
mass functions, and examine the extent to which these may be
biased. Subsequent papers on other samples, in particular a sample
selected at optical wavelengths, will allow us to refine these
estimates. We will also begin to address the other questions raised in this
section.

Section 2 describes the observations and the methods used to reduce
the data. In Section 3 we present the results from the {\em IRAS\/}-selected
sample, in Section 4 we discuss the submillimetre
luminosity and dust mass functions while in Section 5 we compare our results
with other data (gas masses, optical properties) taken from the
literature.\\

We assume a Hubble constant of 75 km s$^{-1}$ Mpc$^{-1}$ throughout.


%% file: obs.tex

\section{OBSERVATIONS AND DATA REDUCTION}
\subsection{The Sample}

The {\em IRAS\/}-selected sample was taken from the revised Bright
Galaxy Sample (BGS), which is complete to a flux limit of
$S_{60}>5.24$ Jy at all $\mid b \mid>30^\circ$ and $\delta >
-30^\circ$ (Soifer et al. 1989). We observed a subset of this sample
with SCUBA, consisting of all galaxies with declination from
$-10^\circ <\delta< 50^\circ$ and with velocity $>$ 1900 km s$^{-1}$,
a limit imposed to try to ensure the galaxies fitted within the SCUBA
field of view. The SCUBA sample covers an area of $\approx$ 10,400
sq$^\circ$ and contains 104 objects. Many of these are interacting
pairs and although most were resolved by SCUBA they were not by {\em
IRAS\/}, even with subsequent HIRES processing (Surace et
al. 1993). Table~\ref{fluxT} lists the sample.

\subsection{Observations}

We observed the galaxies between July 1997 and September 1998 using
the SCUBA bolometer array at the 15-m James Clerk Maxwell Telescope
(JCMT) on Mauna Kea, Hawaii. SCUBA has two arrays of 37 and 91
bolometers for operation at long (850$\mu$m) and short (450$\mu$m)
wavelengths respectively. They operate simultaneously with a field of
view of $\sim$ 2.3 arcmins (slightly smaller at 450$\mu$m). The arrays
are cooled to 0.1 K and have typical sensitivities (NEFDs) of 90 mJy
Hz$^{-1/2}$ at 850$\mu$m and 700 mJy Hz$^{-1/2}$ at 450$\mu$m (Holland
et al. 1999). Beam sizes were measured to be $\sim 15$ arcsec and 8
arcsec at 850 and 450$\mu$m respectively, depending on chop throw and
conditions. We made our observations in `jiggle-map' mode, which is
the most efficient mapping mode for sources smaller than the field of
view. The arrangement of the bolometers is such that the sky is
instantaneously under-sampled and so the secondary mirror is stepped
in a 16-point pattern to `fill in the gaps'. For observations using
both arrays, a 64 offset pattern is required to fully sample the sky as
the spacing/size of the feedhorns is larger for the long wavelength
array. Generally we used the 64-point jiggle, except in some cases
where it was clear that the atmospheric conditions were too poor to
obtain useful 450$\mu$m data, when we instead used the 16-point jiggle
to give better sky cancellation.

The telescope has a chopping secondary mirror operating at 7.8 Hz
which provides cancellation of rapid sky variations. To compensate for
linear sky gradients (the effect of a gradual increase or decrease in
sky brightness), the telescope is nodded to the `off' position when in
`jiggle' mode. We used a chop throw of 120 arcsec in azimuth in all
observations, except for galaxies with nearby companions, when we
chose a chop direction which avoided the second galaxy. Skydips were
performed regularly to measure the zenith opacity $\tau_{850}$. This varied
over the course of the observations resulting in some data being taken
in excellent conditions with $\tau_{850}< 0.2$ while others done in
`back-up' mode had $\tau_{850}\sim$ 0.5--0.6. Due to the large range
of observing conditions, only about one third of the galaxies have
useful 450$\mu$m data and, in fact, we will not consider the 450$\mu$m
data further in this paper although they will be presented in another
paper at a later date.  We checked the pointing regularly
and found it generally to be good to $\sim 2$ arcsec. Due to the
uncertainty of the {\em IRAS\/} positions, we centred our observations
on the optical coordinates taken from the {\em Digitised Sky Survey}
(DSS)\footnote{The Digitised Sky Surveys were produced at the Space
Telescope Science Institute under U.S. Government grant NAGW-2166. The
images of these surveys are based on photographic data obtained using
the Oschin Schmidt Telescope on Palomar Mountain and the U.K. Schmidt
Telescope. The plates were processed into the present digital form
with the permission of these institutions.}.

Integration times differed depending on conditions and source
strength. For bright sources in good weather we typically used 6
integrations (15 mins), but many sources had to be observed on more
than one occasion due to poor S/N or because the galaxy fell on noisy
bolometers. In total we integrated for $\sim 44$ hours, mostly in
weather band 3 ($0.3\leq\tau_{850}\leq0.5$). We calibrated our data by
making jiggle maps of planets (Uranus and Mars) and, when they were
unavailable, of the secondary calibrators CRL 618 and HL Tau. Planet
fluxes were taken from the JCMT {\sc fluxes} program and we assumed
that CRL 618 and HL Tau had fluxes of 4.56 Jy beam$^{-1}$ and 2.32 Jy
beam$^{-1}$ respectively. The calibration error at $850 \mu$m was
taken to be 10 per cent (except for 7 galaxies observed in very poor
conditions where 15 per cent was used instead. These are denoted by a
* in Table~\ref{fluxT}.). This calibration error is included in the
flux error quoted in Table~\ref{fluxT}.

\subsection{Data reduction}

We reduced the data using the standard {\sc surf} package (Jenness \& Lightfoot
1998). This consisted of flat-fielding the data after the on-off
positions had been subtracted and then correcting for atmospheric
extinction using opacities derived from skydip measurements. Any noisy
bolometers were flagged as `bad' and large spikes removed.

\input{table1}

\vspace{2mm}
Except on the most dry and stable nights a residual sky noise was seen
which was correlated across all the pixels. In most cases we removed
this using the {\sc surf} task {\sc remsky}, which uses bolometers
chosen by the user to define a sky region, and then subtracts this sky
level from the other bolometers. The bolometers used to measure the
`sky' were chosen to be those away from the source emission using a
rough submillimetre map and optical information as a guide. After sky
noise removal, the data was despiked and regridded to form an image on
1 arcsec pixels. If an object had more than one dataset, a co-add was
made in which each map was weighted as the inverse square of its
measured noise.

The nodding and chopping should ideally leave a background with zero
mean and some correlated noise superimposed on top. In reality
the background level was not zero and could, in bad conditions, become
quite large (either positive or negative). Often this residual sky
level would vary linearly across the array giving a `tilted sky
plane', the {\sc surf} task {\sc remsky}, which was designed to remove
the sky noise is rather simplistic and cannot remove these sky planes
(Jenness, Lightfoot \& Holland 1998). Due to the short integration
times of most of our objects and the poor conditions in which many
were observed, this problem of sky planes became the limiting factor
in obtaining sufficient S/N in many observations. To this end a
program was written to remove spatially varying sky backgrounds from
the array as well as the temporal `sky noise' dealt with by {\sc
remsky}.

\begin{figure*}
 \vspace{7cm}
 \includegraphics{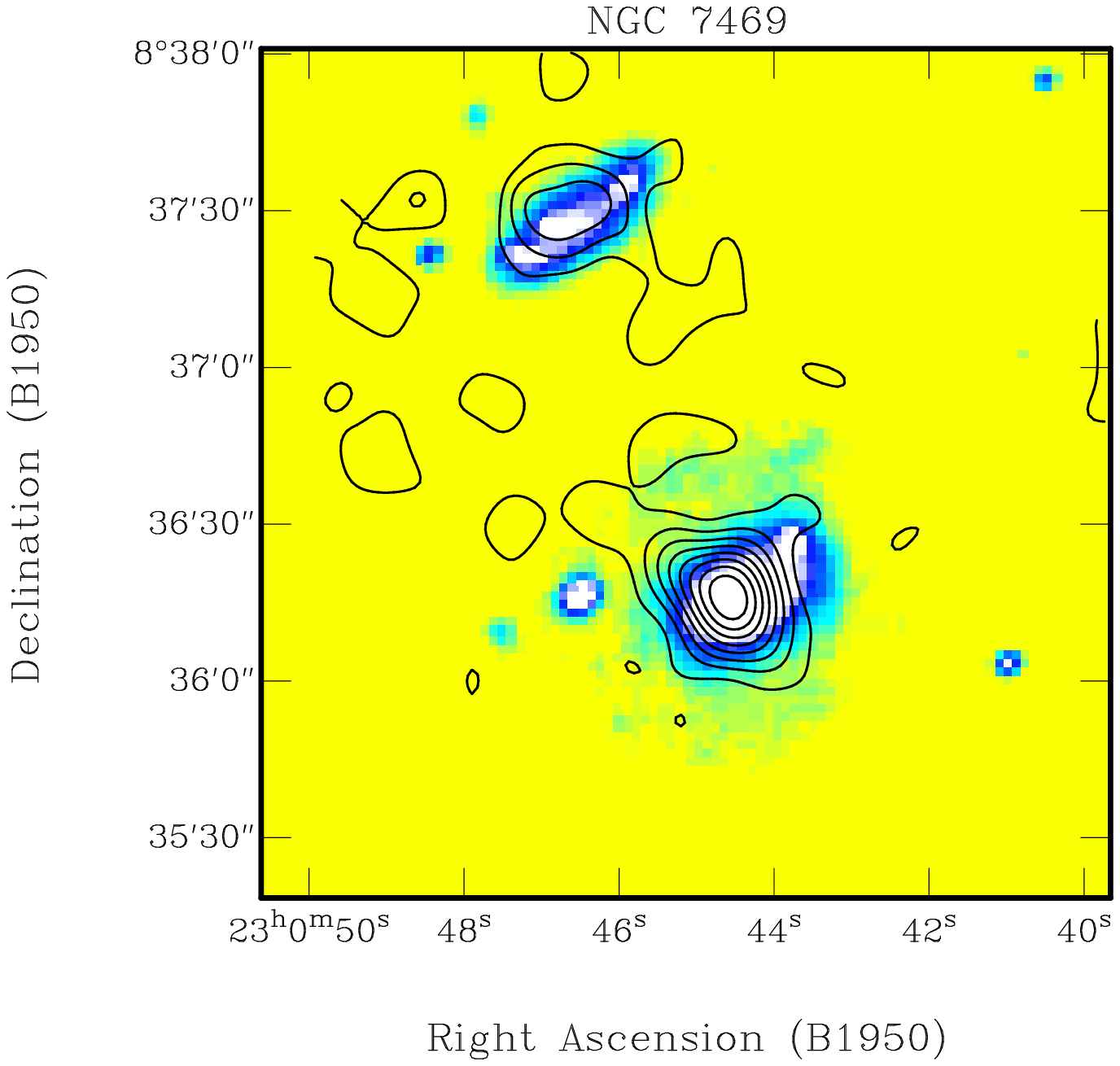}
 \includegraphics{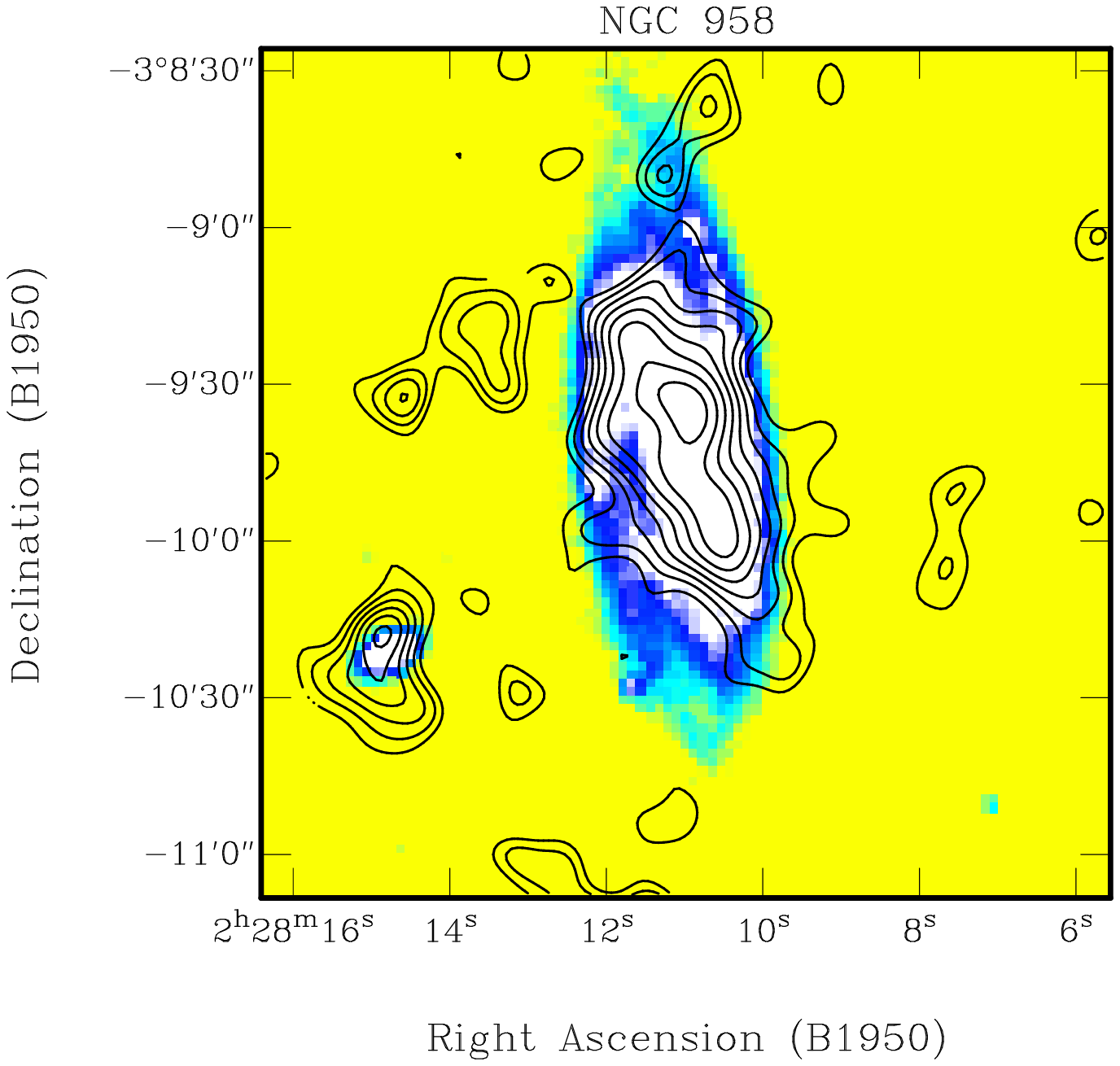}
 \caption{\label{jmapF}a) SCUBA contours ($2, 4, 6 ... \sigma$) overlaid on an optical DSS
image for the Seyfert 1 galaxy NGC 7469 (lower), and its companion IC
5283. There appears to be extended emission between the two galaxies.
b) SCUBA contours ($2,3,4 ... \sigma$) overlaid over the optical DSS image for
NGC 958. The companion satellite $\sim 1$ arcmin to the east is a
surprise detection ($\sim 35$ mJy) as it has no 1.49 GHz flux.}
\end{figure*}

Two fairly typical 850$\mu$m images are shown in Figure~\ref{jmapF} along
with optical images from the DSS. The SCUBA
images clearly contain large amounts of structural information, which
we will delay considering until a later paper. Here we will simply
consider the 850$\mu$m flux measurements.

\input{table2}

\subsection{Flux measurement and error analysis}

We used the submillimetre images to choose an aperture over which to
integrate the flux, trying to ensure that it contained all the flux
associated with the galaxy and as little of the surrounding sky as possible.\\

The noise /error on the flux measurement has two components.\\

i) The error in subtracting the background sky level which, despite
our best attempts at removing it, often had a non-zero value that
varied over the image. A number of smaller apertures were placed on
regions containing no source emission and used to estimate the mean
sky level. The standard deviation between the sky levels in each
aperture was then used to find the error in the mean sky as
$\sigma_{ms} = S.D./\sqrt{\rm{No.\, aps}}$, where `No. aps' is the number of
sky apertures used to determine the mean sky value. The noise in the
object aperture due to this sky error is then
$\sigma_{sky}=\sigma_{ms}N_{ap}$ where $N_{ap}$ is the number of
pixels in the object aperture.\\

ii)  Shot/Poisson noise from pixel to pixel variations within the sky
aperture, which would usually be determined from the standard
deviation of the pixels within the sky aperture and contribute to the
flux error as $\sigma_{shot}=\sigma_{pix}\,\sqrt{N_{ap}}$. In ordinary
CCD astronomy this would be the only error estimated and we will refer
to it as the `theoretical error'. It is actually quite a large
underestimate of the true shot noise because the noise is correlated
across pixels. This is because, unlike a CCD image, the pixels of a
SCUBA image do not contain independent samples of the sky, i.e. two
neighbouring pixels on a SCUBA image are constructed from overlapping
sets of bolometer measurements, (also the pixels on the SCUBA maps are
`artificial' as they are smaller than the size of the beam). To
quantify this effect, simulations of blank SCUBA fields were made by
replacing the data for each bolometer in an observation by the output
of a Gaussian random number generator. A large number of artificial
maps were made in this way, and on each map the noise in an aperture
of given size was measured in the traditional way using
$\sigma_{shot}=\sigma_{pix}\,\sqrt{N_{ap}}$.  This was compared with
the standard deviation of all the fluxes (in that size aperture)
between the different maps, deemed to be the `real shot noise'. For the
range of aperture sizes actually used in measuring fluxes, the ratio
of real to predicted noise was $\sim 8$. The total error for each flux
was calculated from\\

\[
\sigma_{tot}=(\sigma_{cal}^{2} + \sigma_{sky}^{2} +
\sigma_{shot}^{2})^{1/2}
\]

where $\sigma_{cal}\approx 10\%$

\[
\sigma_{sky} = \sigma_{ms}N_{ap}
\]

\[
\sigma_{shot} = 8\,\sigma_{pix}\,\sqrt{N_{ap}}
\]

To convert the aperture flux in volts to Janskys, a measurement of the
calibrator flux for that night was made using the same aperture as was
used for the object. The orientation of the aperture relative to the
chop throw was also translated to the calibrator map (this did have a
significant effect for more elliptical apertures).

Fluxes were found to have total errors $\sigma_{tot}$ in the range
10--25 per cent at 850$\mu$m and these are listed in Table~\ref{fluxT}. All
objects were detected at $> 3 \sigma$ except for IR 0857+39, which was
only detected at $2.4 \sigma$, although the flux is in good agreement
with that measured by Rigopoulou et al. (1996). Where the system is
interacting and not resolved by {\em IRAS\/}, the 850$\mu$m fluxes
given are for both galaxies combined. For those paired galaxies which
were resolved by SCUBA, the individual fluxes are listed in
Table~\ref{pfluxT}.

A few objects were detected with good signal to noise on more
than one occasion and these could be used to keep check on the
consistency of our flux measuring and calibration procedures. These
objects and their relative flux errors are detailed in
Table~\ref{calT} and it can be seen that the differences in measured
fluxes are within the errors calculated in the manner described above.

\input{table3}


%% file: table1.tex

\onecolumn
\begin{longtable}{lccccccccccc} 
\caption{\label{fluxT}Flux densities and SED parameters}\\
\hline
\\[-2.5ex]
\multicolumn{1}{c}{(1)}&\multicolumn{1}{c}{(2)}&\multicolumn{1}{c}{(3)}&\multicolumn{1}{c}{(4)}&\multicolumn{1}{c}{(5)}&\multicolumn{1}{c}{(6)}&\multicolumn{1}{c}{(7)}&\multicolumn{1}{c}{(8)}&\multicolumn{1}{c}{(9)}&\multicolumn{1}{c}{(10)}&\multicolumn{1}{c}{(11)}&\multicolumn{1}{c}{(12)}\\
\\[-1ex]
\multicolumn{1}{c}{Name}&\multicolumn{1}{c}{R.A.}&\multicolumn{1}{c}{Decl.}&\multicolumn{1}{c}{$cz$}&\multicolumn{1}{c}{$S_{60}$}&\multicolumn{1}{c}{$S_{100}$}&\multicolumn{1}{c}{$S_{850}$}&\multicolumn{1}{c}{$\sigma_{850}$}&\multicolumn{1}{c}{$T_{\rm{d}}$}&\multicolumn{1}{c}{$\sigma(T_{\rm{d}})$}&\multicolumn{1}{c}{$\beta$}&\multicolumn{1}{c}{$\sigma(\beta)$}\\
\multicolumn{1}{c}{}&\multicolumn{1}{c}{(J2000)}&\multicolumn{1}{c}{(J2000)}&\multicolumn{1}{c}{(km
s$^{-1}$)}&\multicolumn{1}{c}{(Jy)}&\multicolumn{1}{c}{(Jy)}&\multicolumn{1}{c}{(mJy)}&\multicolumn{1}{c}{(mJy)}&\multicolumn{1}{c}{(K)}&\multicolumn{1}{c}{(K)}&\multicolumn{1}{c}{}&\multicolumn{1}{c}{}\\
\\[-2.5ex]
\hline
\\[-2.5ex]
\endfirsthead
\caption{... continued}\\
\hline
\\[-2.5ex]
\multicolumn{1}{c}{(1)}&\multicolumn{1}{c}{(2)}&\multicolumn{1}{c}{(3)}&\multicolumn{1}{c}{(4)}&\multicolumn{1}{c}{(5)}&\multicolumn{1}{c}{(6)}&\multicolumn{1}{c}{(7)}&\multicolumn{1}{c}{(8)}&\multicolumn{1}{c}{(9)}&\multicolumn{1}{c}{(10)}&\multicolumn{1}{c}{(11)}&\multicolumn{1}{c}{(12)}\\
\\[-1ex]
\multicolumn{1}{c}{Name}&\multicolumn{1}{c}{R.A.}&\multicolumn{1}{c}{Decl.}&\multicolumn{1}{c}{$cz$}&\multicolumn{1}{c}{$S_{60}$}&\multicolumn{1}{c}{$S_{100}$}&\multicolumn{1}{c}{$S_{850}$}&\multicolumn{1}{c}{$\sigma_{850}$}&\multicolumn{1}{c}{$T_{\rm{d}}$}&\multicolumn{1}{c}{$\sigma(T_{\rm{d}})$}&\multicolumn{1}{c}{$\beta$}&\multicolumn{1}{c}{$\sigma(\beta)$}\\
\multicolumn{1}{c}{}&\multicolumn{1}{c}{(J2000)}&\multicolumn{1}{c}{(J2000)}&\multicolumn{1}{c}{(km
s$^{-1}$)}&\multicolumn{1}{c}{(Jy)}&\multicolumn{1}{c}{(Jy)}&\multicolumn{1}{c}{(mJy)}&\multicolumn{1}{c}{(mJy)}&\multicolumn{1}{c}{(K)}&\multicolumn{1}{c}{(K)}&\multicolumn{1}{c}{}&\multicolumn{1}{c}{}\\
\\[-2.5ex]
\hline
\\[-2.5ex]
\endhead
\\[-2.5ex]
\hline
\endfoot
\\[-2.5ex]   
\hline 
\endlastfoot
NGC 23 & 0 09 53.4 & $+$25 55 26 & 4565 & 8.77 & 14.96 & 144 & 25 &
34.4 & 3.0 & 1.3 & 0.2\\ 
UGC 556 & 0 54 50.3 & $+$29 14 47 & 4629 & 5.36 & 9.99 & 79 & 15
& 32.0 & 2.6 & 1.5 & 0.2\\ 
NGC 470 & 1 19 44.8 & $+$03 24 35 & 2374 & 7.09 & 12.01 &
213 & 30 & 38.0 & 3.0 & 0.9 & 0.2\\ 
MCG+02-04-025 & 1 20 02.7 & $+$14 21 43 & 9362 & 10.72 & 9.60 & 39 & 8
& 44.6 & 4.6 & 1.4 & 0.2 \\ 
UGC 903 & 1 21 47.9 & $+$17 35 34 & 2518 & 7.91 & 14.58 & 178 & 26 &
34.4 & 3.1 & 1.2 & 0.2\\
NGC 520 & 1 24 34.9 & $+$03 47 31 & 2281 & 31.55 & 46.56 & 325 & 50 &
36.2 & 3.1 & 1.4 & 0.2\\
III ZW 035 & 1 44 30.5 & $+$17 06 08 & 8215 & 11.86 & 13.75 & 76 & 15
& 39.2 & 3.8 & 1.4 & 0.2\\ 
NGC 695 & 1 51 14.3 & $+$22 34 56 & 9735 & 7.61 & 13.80 & 136 & 21 &
33.8 & 2.8 & 1.3 & 0.2\\
NGC 697 & 1 51 17.5 & $+$22 21 30 & 3117 & 5.62 & 16.54 & 221 & 48* &
27.8 & 2.2 & 1.5 & 0.2\\ 
UGC 1351 & 1 52 59.6 & $+$12 42 31 & 4558 & 6.12 & 11.71 & 141 & 21 &
34.4 & 2.8 & 1.2 & 0.2\\ 
UGC 1451 & 1 58 30.0 & $+$25 21 36 & 4916 & 6.75 & 12.20 & 107 & 18 &
33.2 & 2.6 & 1.4 & 0.2\\ 
NGC 772 & 1 59 19.5 & $+$19 00 30 & 2472 & 6.78 & 24.11 & 288 & 43 &
25.4 & 1.7 & 1.7 & 0.2\\
NGC 877 & 2 17 59.5 & $+$14 32 42 & 3913 & 11.76 & 23.34 & 332 & 43 &
34.4 & 2.8 & 1.1 & 0.2\\
NGC 958 & 2 30 42.8 & $-$02 56 23 & 5738 & 5.90 & 14.99 & 262 & 34 &
30.8 & 2.0 & 1.2 & 0.2\\ 
NGC 992 & 2 37 25.5 & $+$21 06 03 & 4141 & 10.96 & 15.63 & 146 & 26 &
38.0 & 3.5 & 1.2 & 0.2\\
UGC 2238 & 2 46 17.5 & $+$13 05 48 & 6436 & 8.16 & 15.22 & 104 & 14 &
31.4 & 2.9 & 1.6 & 0.2\\ 
IR 0243+21 & 2 46 39.2 & $+$21 35 11 & 6987 & 5.50 & 6.25 & 40 & 8 &
41.0 & 4.5 & 1.3 & 0.2\\ 
NGC 1134 & 2 53 41.5 & $+$13 00 53 & 3651 & 8.99 & 16.07 & 242 & 31 &
36.2 & 3.4 & 1.0 & 0.2\\ 
UGC 2369 & 2 54 01.8 & $+$14 58 14 & 9400 & 7.68 & 11.10 & 72 & 13 &
36.2 & 3.4 & 1.4 & 0.2\\
UGC 2403 & 2 55 57.2 & $+$00 41 33 & 4161 & 7.51 & 11.77 & 111 & 18 &
36.8 & 2.9 & 1.2 & 0.2\\ 
NGC 1222 & 3 08 56.8 & $-$02 57 18 & 2452 & 12.86 & 15.15 & 84 & 16 &
39.2 & 4.0 & 1.4 & 0.2\\ 
IR 0335+15 & 3 38 47.1 & $+$15 32 53 & 10600 & 5.77 & 6.53 & 44 & 9 &
42.2 & 4.9 & 1.2 & 0.2\\ 
UGC 2982 & 4 12 22.5 & $+$05 32 51 & 5305 & 8.70 & 17.32 & 176 & 34* &
32.0 & 2.8 & 1.4 & 0.2\\
NGC 1614 & 4 34 00.0 & $-$08 34 45 & 4778 & 33.12 & 36.19 & 219 & 32 &
41.6 & 4.2 & 1.3 & 0.2\\ 
NGC 1667 & 4 48 37.2 & $-$06 19 12 & 4547 & 6.24 & 16.54 & 163 & 22 &
28.4 & 1.7 & 1.6 & 0.2\\  
NGC 2623 & 8 38 24.1 & $+$25 45 16 & 5535 & 25.72 & 27.36 & 91 & 14 & 39.8
& 4.1 & 1.6 & 0.2\\ 
IR 0857+39 & 9 00 24.5 & $+$39 03 55 & 17480 & 7.66 & 5.06 & 17 & 7 &
50.6 & 9.0 & 1.4 & 0.4\\
NGC 2782 & 9 14 05.1 & $+$40 06 49 & 2562 & 9.60 & 14.65 & 237 & 32 &
39.2 & 3.6 & 0.9 & 0.2\\
NGC 2785 & 9 15 15.4 & $+$40 55 03 & 2734 & 9.21 & 16.78 & 201 & 33 &
34.4 & 3.1 & 1.2 & 0.2\\ 
UGC 4881 & 9 15 54.7 & $+$44 19 52 & 11782 & 6.53 & 10.21 & 65 & 13 &
34.4 & 3.1 & 1.5 & 0.2\\ 
NGC 2856 & 9 24 16.2 & $+$49 14 58 & 2638 & 6.15 & 10.28 & 89 & 16 &
35.0 & 2.9 & 1.3 & 0.2\\
MCG+08-18-012 & 9 36 37.2 & $+$48 28 28 & 7777 & 6.39 & 8.83 & 42 & 10 &
35.6 & 3.4 & 1.6 & 0.2\\ 
NGC 2966 & 9 42 11.5 & $+$04 40 23 & 2044 & 5.76 & 8.69 & 136 & 28 &
39.8 & 4.9 & 0.9 & 0.2\\
NGC 2990 & 9 46 17.2 & $+$05 42 33 & 3088 & 5.49 & 10.16 & 110 & 19 &
33.8 & 3.2 & 1.3 & 0.2\\
IC 563/4 $^p$ & 9 46 20.7 & $+$03 03 31 & 6020 & 5.35 & 12.43 & 228 & 35
& 32.0 & 3.0 & 1.1 & 0.2\\
UGC 5376 & 10 00 26.8 & $+$03 22 26 & 2050 & 5.94 & 11.49 & 148 & 23 &
33.8 & 2.9 & 1.2 & 0.2\\
NGC 3094 & 10 01 26.0 & $+$15 46 13 & 2404 & 11.54 & 15.10 & 152 & 31 &
40.4 & 3.8 & 1.1 & 0.2\\
NGC 3110 & 10 04 02.0 & $-$06 28 31 & 5048 & 11.68 & 23.16 & 188 & 28
& 32.0 & 2.3 & 1.5 & 0.2\\
IR 1017+08 & 10 20 00.2 & $+$08 13 34 & 14390 & 6.08 & 5.97 & 36 & 6 &
44.0 & 4.3 & 1.2 & 0.2\\
NGC 3221 & 10 22 20.1 & $+$21 34 09 & 4110 & 7.44 & 19.56 & 253 & 37 &
29.6 & 2.2 & 1.4 & 0.2\\
NGC 3367 & 10 46 34.6 & $+$13 45 03 & 3037 & 6.06 & 12.49 & 132 & 21 &
31.4 & 2.6 & 1.4 & 0.2\\
IR 1056+24 & 10 59 18.2 & $+$24 32 34 & 12912 & 12.53 & 16.06 & 61 & 13
& 35.6 & 3.3 & 1.7 & 0.2\\
ARP 148 & 11 03 54.0 & $+$40 50 59 & 10350 & 6.95 & 10.99 & 92 & 20 & 35.6
& 3.0 & 1.3 & 0.2\\
NGC 3583 & 11 14 12.2 & $+$48 18 56 &2136 & 7.18 & 19.50 & 185 &
31 & 28.4 & 2.2 & 1.6 & 0.2\\
MCG+00-29-023 & 11 21 10.9 & $-$02 59 13 & 7646 & 5.40 & 8.87 & 84 &
13 & 35.0 & 3.6 & 1.3 & 0.2\\
UGC 6436 & 11 25 46.6 & $+$14 40 26 & 10243 & 5.60 & 9.80 & 106 & 18 &
35.0 & 3.2 & 1.2 & 0.2\\
NGC 3994/5 $^p$ & 11 57 39.4 & $+$32 17 08 & 3170 & 8.26 & 16.94 & 232 &
31 & 33.2 & 2.5 & 1.2 & 0.2\\
NGC 4045 & 12 02 42.2 & $+$01 58 37 & 1981 & 6.50 & 13.57 & 142 & 25 &
31.4 & 2.6 & 1.4 & 0.2\\
IR 1211+03 & 12 13 46.1 & $+$02 48 40 & 21703 & 8.39 & 9.10 & 49 & 10
& 42.2 & 5.2 & 1.3 & 0.2\\
NGC 4273 & 12 19 56.1 & $+$05 20 37 & 2378 & 10.52 & 21.02 & 324 & 46
& 33.8 & 3.2 & 1.1 & 0.2\\
IR 1222-06 & 12 25 03.9 & $-$06 40 53 & 7902 & 5.79 & 7.53 & 74 & 15 &
40.4 & 4.9 & 1.1 & 0.2\\
NGC 4418 & 12 26 54.7 & $-$00 52 39 & 2179 & 42.32 & 30.76 & 255 & 37
& 55.4 & 8.0 & 0.9 & 0.2\\
NGC 4433 & 12 27 38.7 & $-$08 16 42 & 3000 & 14.15 & 22.42 & 220 & 47
& 36.8 & 3.5 & 1.2 & 0.2\\
NGC 4793 & 12 54 41.1 & $+$28 56 21 & 2484 & 12.49 & 27.99 & 258 & 42
& 30.2 & 2.5 & 1.5 & 0.2\\
NGC 4922 & 13 01 25.3 & $+$29 18 49 & 7071 & 6.20 & 7.30 & 53 & 12 &
41.6 & 4.4 & 1.2 & 0.2\\
NGC 5020 & 13 12 39.8 & $+$12 35 58 & 3362 & 5.39 & 10.01 & 206 & 34 &
36.2 & 3.1 & 0.9 & 0.2\\
IC 860 & 13 15 03.5 & $+$24 37 08 & 3347 & 17.66 & 17.66 & 118 & 20 &
43.4 & 4.4 & 1.2 & 0.2\\
UGC 8387 & 13 20 35.3 & $+$34 08 22 & 7000 & 13.69 & 24.90 & 113 & 15 &
30.8 & 2.4 & 1.8 & 0.2\\
NGC 5104 & 13 21 23.2 & $+$00 20 32 & 5578 & 6.69 & 12.77 & 91 & 20* &
31.4 & 2.9 & 1.6 & 0.2\\
NGC 5256 & 13 38 17.5 & $+$48 16 36 & 8353 & 7.19 & 10.35 & 82 & 17 &
36.8 & 4.6 & 1.3 & 0.2\\
NGC 5257/8 $^p$ & 13 39 55.5 & $+$00 50 09 & 6770 & 10.68 & 20.80 &
283 & 39* & 33.2 & 2.4 & 1.2 & 0.2\\
UGC 8739 & 13 49 14.0 & $+$35 15 24 & 5032 & 5.90 & 14.32 & 187 & 27 &
30.8 & 2.4 & 1.3 & 0.2\\
NGC 5371 & 13 55 39.0 & $+$40 27 31 & 2553 & 5.40 & 18.16 & 136 & 24 &
25.4 & 1.5 & 1.9 & 0.2\\
NGC 5394/5 $^p$ & 13 58 38.1 & $+$37 26 27 & 3470 & 9.07 & 21.51 & 269
& 36 & 30.2 & 2.6 & 1.4 & 0.2\\
NGC 5433 & 14 02 36.0 & $+$32 30 38 & 4354 & 6.34 & 11.26 & 163 & 28 &
35.6 & 3.4 & 1.1 & 0.2\\
NGC 5426/7 $^p$ & 14 03 26.0 & $-$06 01 50 & 2678 & 9.93 & 24.81 & 548
& 86* & 32.0 & 2.8 & 1.0 & 0.2\\
ZW 247.020 & 14 19 43.3 & $+$49 14 12 & 7666 & 5.91 & 8.25 & 36 & 8 &
34.4 & 3.4 & 1.7 & 0.2\\
NGC 5600 & 14 23 49.2 & $+$14 38 22 & 2319 & 5.35 & 11.46 & 112 & 22 &
31.4 & 2.6 & 1.4 & 0.2\\
NGC 5653 & 14 30 10.2 & $+$31 12 56 & 3562 & 10.27 & 21.86 & 205 & 32
& 30.8 & 2.0 & 1.5 & 0.2\\
NGC 5665 & 14 32 25.9 & $+$08 04 47 & 2228 & 6.18 & 11.67 & 157 & 27 &
35.0 & 3.4 & 1.1 & 0.2\\
NGC 5676 & 14 32 46.7 & $+$49 27 29 & 2114 & 12.00 & 29.78 & 481 & 69
& 30.8 & 1.8 & 1.2 & 0.2\\
NGC 5713 & 14 40 11.4 & $-$00 17 26 & 1900 & 20.69 & 36.27 & 359 & 44
& 34.4 & 3.1 & 1.3 & 0.2\\
UGC 9618 $^p$ & 14 57 00.5 & $+$24 36 42 & 9900 & 6.68 & 14.54 & 215 &
31 & 32.0 & 2.6 & 1.2 & 0.2\\
NGC 5792 & 14 58 22.7 & $-$01 05 29 & 1924 & 9.45 & 18.31 & 383 & 56 &
35.6 & 3.2 & 0.9 & 0.2\\
ZW 049.057 & 15 13 13.1 & $+$07 13 31 & 3927 & 21.06 & 29.88 & 200 &
27 & 36.2 & 3.2 & 1.4 & 0.2\\
NGC 5900 & 15 15 05.2 & $+$42 12 36 & 2511 & 7.36 & 16.69 & 179 & 23 &
30.8 & 2.6 & 1.4 & 0.2\\
1 Zw 107 & 15 18 06.1 & $+$42 44 45 & 11946 & 9.15 & 10.04 & 60 & 14 &
41.6 & 4.6 & 1.3 & 0.2\\
NGC 5929/30 $^p$ & 15 26 06.3 & $+$41 40 24 & 2617 & 9.14 & 13.69 &
119 & 22 & 36.2 & 3.1 & 1.3 & 0.2\\
IR 1525+36 & 15 26 59.4 & $+$35 58 37 & 16009 & 7.20 & 5.78 & 43 & 10
& 50.6 & 6.9 & 1.0 & 0.2\\
NGC 5936 & 15 30 00.9 & $+$12 59 21 & 4004 & 8.56 & 16.84 & 152 & 28 &
32.6 & 2.9 & 1.4 & 0.2\\
NGC 5937 & 15 30 46.1 & $-$02 49 46 & 2758 & 10.23 & 20.56 & 247 & 39
& 32.6 & 2.7 & 1.3 & 0.2\\
NGC 5953/4 $^p$ & 15 34 33.7 & $+$15 11 50 & 1960 &11.55 & 19.50 & 306
& 31 & 36.8 & 4.1 & 1.0 & 0.2\\
ARP 220 & 15 34 57.2 & $+$23 30 11 & 5452 & 103.33 & 113.95 & 832 & 86
& 42.2 & 5.7 & 1.2 & 0.2\\
IR 1533-05 & 15 36 11.7 & $-$05 23 52 & 8186 & 5.25 & 8.96 & 79 & 15 &
35.0 & 3.0 & 1.3 & 0.2\\
NGC 5962 & 15 36 32.0 & $+$16 36 22 & 1963 & 8.99 & 20.79 & 317 & 37 &
32.0 & 2.7 & 1.2 & 0.2\\
NGC 5990 & 15 46 16.4 & $+$02 24 55 & 3839 & 9.20 & 15.46 & 110 & 23 &
33.2 & 2.7 & 1.5 & 0.2\\
NGC 6052 & 16 05 13.0 & $+$20 32 34 & 4712 & 6.46 & 10.18 & 95 & 15 &
35.6 & 3.5 & 1.3 & 0.2\\
MCG+01-42-008 & 16 30 56.5 & $+$04 04 58 & 7342 & 7.38 & 12.48 & 95 &
18 & 34.4 & 2.8 & 1.4 & 0.2\\
NGC 6181 & 16 32 21.2 & $+$19 49 30 & 2379 & 9.35 & 21.00 & 228 & 37 &
30.8 & 2.6 & 1.4 & 0.2\\
NGC 7448 & 23 00 03.6 & $+$15 58 56 & 2192 & 8.32 & 17.08 & 193 & 32 &
32.6 & 2.8 & 1.3 & 0.2\\
NGC 7469 $^p$ & 23 03 16.4 & $+$08 52 50 & 4892 & 27.68 & 34.91 & 264 & 30 &
40.4 & 3.9 & 1.2 & 0.2\\
NGC 7479 & 23 04 56.5 & $+$12 19 19 & 2381 & 15.35 & 24.60 & 335 & 56
& 38.0 & 3.8 & 1.0 & 0.2\\
ZW 453.062 & 23 04 56.5 & $+$19 33 08 & 7524 & 7.06 & 10.39 & 69 & 14
& 36.2 & 3.5 & 1.4 & 0.2\\
NGC 7541 & 23 14 43.4 & $+$04 32 04 & 2665 & 20.59 & 40.63 & 427 & 60
& 33.2 & 2.4 & 1.3 & 0.2\\
ZW 475.056 & 23 16 00.6 & $+$25 33 25 & 8197 & 8.75 & 11.64 & 77 & 15
& 37.4 & 3.8 & 1.4 & 0.2\\
NGC 7591 & 23 18 16.3 & $+$06 35 10 & 4956 & 7.82 & 13.52 & 135 & 21 & 34.4
& 2.8 & 1.3 & 0.2\\
NGC 7592 & 23 18 22.1 & $-$04 24 58 & 7350 & 8.02 & 10.50 & 108 & 19 &
39.8 & 3.8 & 1.1 & 0.2\\
NGC 7674 & 23 27 56.7 & $+$08 46 44 & 8713 & 5.28 & 7.91 & 108 & 20 &
38.6 & 4.2 & 1.0 & 0.2\\
NGC 7678 & 23 28 26.2 & $+$22 25 02 & 3489 & 7.01 & 14.84 & 195 & 26 &
33.2 & 2.7 & 1.2 & 0.2\\
NGC 7679 & 23 28 46.7 & $+$03 30 41 & 5138 & 7.28 & 10.65 & 93 & 15 &
38.0 & 3.6 & 1.2 & 0.2\\
NGC 7714 & 23 36 14.1 & $+$02 09 18 & 2798 & 10.52 & 11.66 & 72 & 13 &
41.0 & 6.9 & 1.3 & 0.2\\
NGC 7771 & 23 51 24.9 & $+$20 06 42 & 4256 & 20.46 & 37.42 & 377 & 42
& 33.8 & 3.0 & 1.3 & 0.2\\
MRK 331 & 23 51 26.8 & $+$20 35 10 & 5541 & 17.32 & 20.86 & 132 & 25 &
40.4 & 4.3 & 1.3 & 0.2\\
UGC 12914/5 $^p$ & 00 01 39.6 & $+$23 29 33 & 4350 & 6.27 & 13.40 &
191 & 30 & 32.6 & 2.6 & 1.2 & 0.2\\

\end{longtable}
\twocolumn
{\bf Notes to Table~\ref{fluxT}:}\\
(1) Most commonly used name taken from the IRAS BGS (2) Right
ascension J2000 epoch (3) Declination J2000 epoch (4) Recession
velocity taken from NED\footnote{The NASA/IPAC Extragalactic Database
(NED) is operated by the Jet Propulsion Laboratory, California
Institute of Technology, under contract with the National Aeronautics
and Space Administration.} (5) 60$\mu$m flux from Soifer et al. 1989
(7) 850$\mu$m flux (this work) (8) error on 850$\mu$m flux, calculated
in the manner described in Section 2.4 and inclusive of a 10\%
calibration uncertainty. A * indicates that a 15\% calibration
uncertainty was used. (9) Dust temperature derived from a single
component fit to the 60, 100 and 850$\mu$ data points as described in
Section 3.1 (10) Statistical uncertainty in the dust temperature as
described in Section 3.1 (11) Emissivity index derived fron the single
component fit, described in Section 3.1 (12) Statistical uncertainty
in the emissivity index as described in Section 3.1 $^p$ indicates a
close or interacting pair which was resolved by SCUBA. Individual
fluxes are listed in Table~\ref{pfluxT}.


%% file: table2.tex

\begin{table*}
\centering
\caption{\label{pfluxT}Fluxes for pairs resolved by SCUBA}
\begin{tabular}{lcccrrcccccc}
\\[-2ex] 
\hline
\\[-2.5ex]
\multicolumn{1}{c}{Name}&\multicolumn{1}{c}{R.A.}&\multicolumn{1}{c}{Decl.}&\multicolumn{1}{c}{$cz$}&\multicolumn{1}{c}{$S_{60}$}&\multicolumn{1}{c}{$S_{100}$}&\multicolumn{1}{c}{$S_{850}$}&\multicolumn{1}{c}{$\sigma_{850}$}&\multicolumn{1}{c}{$T_{\rm{d}}$}&\multicolumn{1}{c}{$\sigma(T_{\rm{d}})$}&\multicolumn{1}{c}{$\beta$}&\multicolumn{1}{c}{$\sigma(\beta)$}\\
\multicolumn{1}{c}{}&\multicolumn{1}{c}{(J2000)}&\multicolumn{1}{c}{(J2000)}&\multicolumn{1}{c}{(km
s$^{-1}$)}&\multicolumn{1}{c}{(Jy)}&\multicolumn{1}{c}{(Jy)}&\multicolumn{1}{c}{(mJy)}&\multicolumn{1}{c}{(mJy)}&\multicolumn{1}{c}{(K)}&\multicolumn{1}{c}{(K)}&\multicolumn{1}{c}{}&\multicolumn{1}{c}{}\\
\\[-2.5ex]
\hline
\\[-2.5ex]
IC 563 & 9 46 20.3 & $+$03 02 46 & 6093 & $^h$2.90 & $^h$5.87 & 103 & 24 & 34.4
& 2.7 & 1.0 & 0.2\\
IC 564 & 9 46 21.0 & $+$03 04 19 & 6026 & $^h$2.46 & $^h$6.56 & 125 &
26 & 30.8 & 2.0 & 1.1 & 0.2\\
\\[-2ex]
NGC 3994 & 11 57 36.9 & $+$32 16 38 & 3096 & $^h$5.00 & $^h$10.31 & 106 & 20 &
32.0 & 2.3 & 1.4 & 0.2\\
NGC 3995 & 11 57 44.0 & $+$32 17 37 & 3254 & $^h$3.27 & $^h$6.63 & 126 & 24 &
34.4 & 3.1 & 1.0 & 0.2\\
\\[-2ex]
NGC 5257 & 13 39 52.1 & $+$00 50 27 & 6798 & U & U & 114 & 23 & ... &...  \\
NGC 5258 & 13 39 57.4 & $+$00 49 51 & 6757 & U & U & 169 & 32 & ... &...  \\
\\[-2ex]
NGC 5394 & 13 58 33.7 & $+$37 27 13 & 3451 &U &U  & 71 & 17 &...  & ... \\
NGC 5395 & 13 58 37.6 & $+$37 25 42 & 3491 &U & U & 198 & 32 &...  &...\\
\\[-2ex]
NGC 5426 & 14 03 24.7 & $-$06 04 13 & 2621 & $^h$3.06 & $^h$8.50 & 287
& 63 & 32.0 & 2.6 & 0.8 & 0.2\\
NGC 5427 & 14 03 26.0 & $-$06 01 50 & 2730 & $^h$6.87 & $^h$16.31 &
261 & 59 & 31.4 & 2.8 & 1.2 & 0.3\\
\\[-2ex] 
UGC 9618 ({\sc n}) & 14 57 00.7 & $+$24 36 58 & 10094 &U & U & 135 & 24 &...  &...  \\
UGC 9618 ({\sc s}) & 14 57 00.0 & $+$24 36 20 & 9776 &U & U & 80 & 20 &...  &... \\
\\[-2ex]
NGC 5929 & 15 26 06.1 & $+$41 40 15 & 2561 & U & U & $\sim$24 & 12
&...&...\\
NGC5930 & 15 26 07.9 & $+$41 40 34 & 2672 & U& U& 95 & 18 &... &...\\
\\[-2ex]
NGC 5953 & 15 34 32.4 & $+$15 11 40 & 1965 &U & U & 182 & 23 &...  &... \\
NGC 5954 & 15 34 35.1 & $+$15 12 04 & 1959 &U &U  & 124 & 21 & ... &... \\
\\[-2ex]
NGC 7469 & 23 03 15.6 & $+$08 52 29 & 4916 &U &U  & 192 & 27 & ... & ... \\
IC 5283 & 23 03 18.2 & $+$08 53 36 & 4894 &  U & U & 72 & 12 &...  &...\\ 
\\[-2ex]
UGC 12914 & 00 01 38.3 & $+$23 29 02 & 4371 &U & U& 131 & 22 &...  & ... \\
UGC 12915 & 00 01 42.1 & $+$23 29 23 & 4336 & U& U& 160 & 21 & ... &...
\\
\\[-2.5ex]
\hline
\\[-2.5ex]
\multicolumn{10}{p{40em}}{\small {\sc notes\/} --  $h$ :- {\it IRAS\/}
fluxes resovled using HIRES (Surace et al 1993),  U :- unresolved by {\it
IRAS\/}. Columns have the same meanings as Table~\ref{fluxT}}
\end{tabular}
\end{table*}

%% file: table3.tex

\parindent=0pt
\begin{table*}
\centering
\caption{\label{calT} Galaxies with repeat flux measurements and
their errors}
\begin{tabular}{lcccccc} 
\\[-2ex]
\hline
\\[-2ex]
\multicolumn{1}{c}{(1)}&\multicolumn{1}{c}{(2)}&\multicolumn{1}{c}{(3)}&\multicolumn{1}{c}{(4)}&\multicolumn{1}{c}{(5)}&\multicolumn{1}{c}{(6)}&\multicolumn{1}{c}{(7)}\\
\\[-2.5ex]
\multicolumn{1}{c}{Object}&\multicolumn{1}{c}{Date}&\multicolumn{1}{c}{Ints.}&\multicolumn{1}{c}{Weather}&\multicolumn{2}{c}{\% error on individual
observations}&\multicolumn{1}{c}{Actual difference}\\
\\[-2ex]
\multicolumn{3}{c}{}&\multicolumn{1}{c}{($\tau_{850}$)}&\multicolumn{1}{c}{Theoretical
noise}&\multicolumn{1}{c}{`True' noise}&\multicolumn{1}{c}{\%}\\
\\[-2ex]
\hline
\\[-2ex]
Arp 220 & 2.7.97 & 5 & 0.2 & 0.3 \hspace{2ex} (10.0)& 2.6 \hspace{2ex} (10.3) & 1.7\\
        & 5.3.98 & 4 & 0.2 & 0.3 \hspace{2ex} (10.0)& 3.0 \hspace{2ex} (10.4) &    \\
\\[-2ex]
NGC 1134 & 7.8.97 & 10 & 0.3 & 0.9 \hspace{2ex} (10.0)& 8.8 \hspace{2ex} (13.3)& 5.0  \\
         & 7.8.97 & 10 & 0.3 & 1.1 \hspace{2ex} (10.1) & 11.6 \hspace{2ex} (15.3)&  \\
\\[-2ex]
UGC 2238 & 14.8.97 & 10 & 0.4--0.5 & 1.9 \hspace{2ex} (10.2)& 15.7 \hspace{2ex} (18.6)& 13.3\\
         & 25.8.97 & 10 & 0.4--0.5 & 1.3 \hspace{2ex} (10.1)& 11.0 \hspace{2ex} (14.9)&   \\
\\[-2ex]
UGC 903  & 14.8.97 & 10 & 0.4--0.5 & 1.2 \hspace{2ex} (10.1)& 10.6 \hspace{2ex} (14.6)& 10.6\\
         & 7.10.98 & 6 & 0.2 & 1.0 \hspace{2ex} (10.0)& 8.3 \hspace{2ex} (13.0)&  \\
\\[-2ex]
\hline
\\[-2ex]
\multicolumn{7}{p{40em}}{\small (1) Galaxy name. (2) Date of each
observation. (3) Number of integrations in observation. (4) Weather in 
terms of $850\mu$m opacity. (5) The `theoretical' noise on the
individual observations, equivalent to shot noise on a CCD. Brackets
give the error inclusive of 10\% calibration uncertainty. (6) The `true' noise on
the individual observations, consisting of $\sigma_{sky}$ and the
`real' $\sigma_{shot}$ as described in Section 2.4, brackets again
give error inclusive of 10\% calibration uncertainty. (7) The
actual percentage difference between the fluxes for both observations. }
\end{tabular}
\end{table*}


%% file: results.tex

\parindent = 1em

\section{RESULTS}

\subsection{Spectral fits}

\begin{figure*}
\vspace{23cm}
\includegraphics{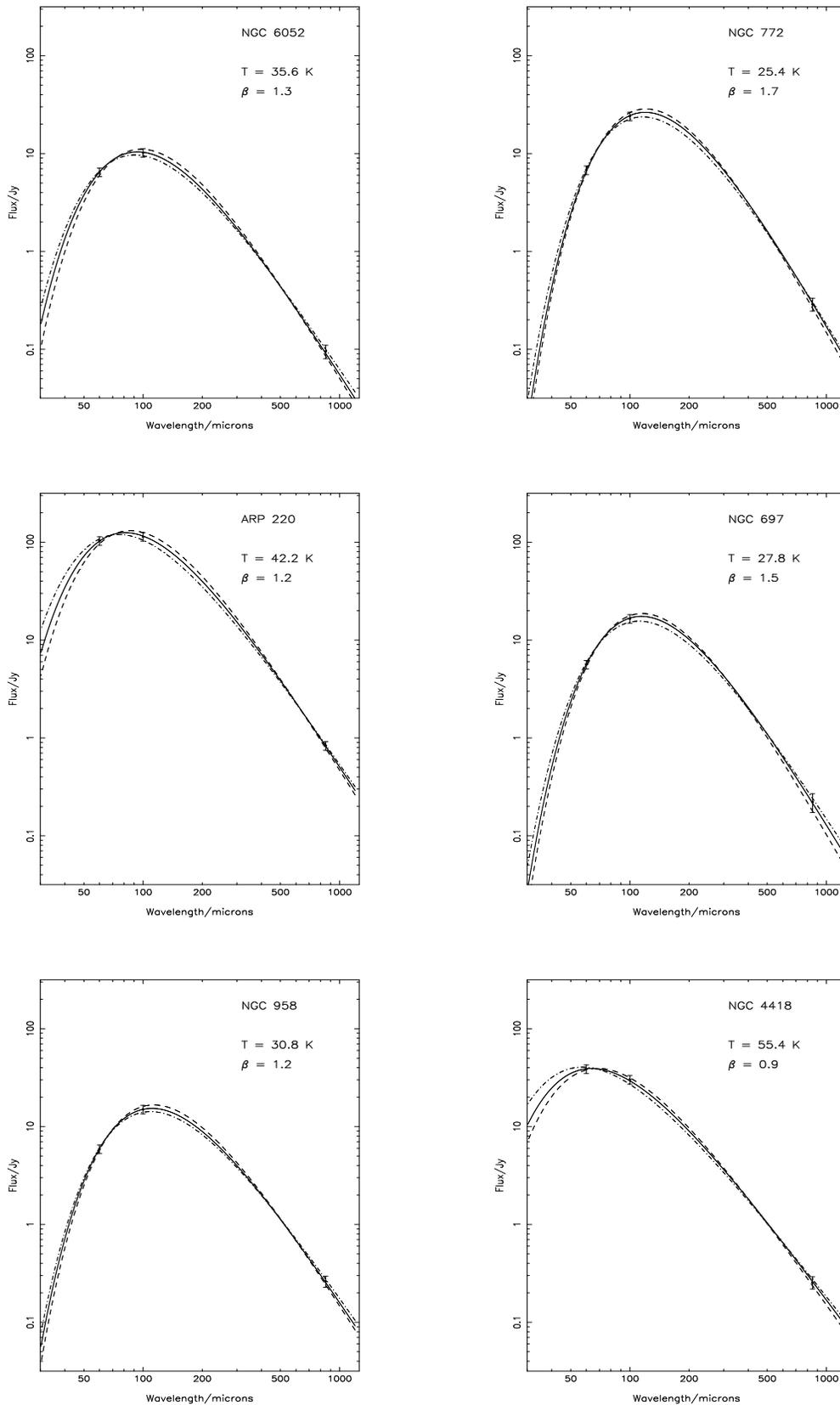}
\caption{\label{sedF} Representative SEDs showing the range of $\beta$ 
and $T_{\rm{d}}$ for the sample. Data points are the 60 and
100$\mu$m {\em IRAS} fluxes and the 850$\mu$m point is from our
survey 10 
. The solid lines are the best-fitting lines relating to the SED
parameters shown. The dashed and dot-dashed lines are the $\pm 1
\sigma$ fits.}
\end{figure*}

We have fitted the {\em IRAS\/} 60, 100 and 850$\mu$m fluxes with a
single-component temperature model by minimising the sum of the
chi-squared residuals. The best fit was found for both the emissivity
index $\beta$ and temperature $T_{\rm{d}}$, with the values given in
Table~\ref{fluxT}. Some examples of fits are shown in Fig.~\ref{sedF}
along with the $1 \sigma$ ranges. The uncertainty of each fitted value
of $T_{\rm{d}}$ and $\beta$ was estimated in the following way. For
each galaxy, a Gaussian random number generator was used to create 100
artificial flux sets from the original fluxes and measurement
errors. These new data sets were then fitted in the same way and the
standard deviation in the new parameters was taken to represent the
uncertainty in the parameters found from the real data set. Average
values and standard deviations (S.D.) for the whole sample (i.e. using
only the best fitting parameters for the real data sets) are
$\overline{T_{\rm{d}}} = 35.6\, \pm\, 4.9$ K and $\overline{\beta} =
1.3\, \pm
\,0.2 $.

\begin{figure}
 \vspace{6cm}
 \includegraphics{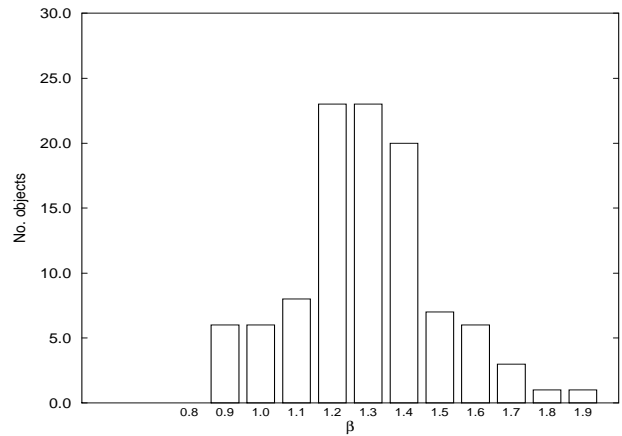}
 \caption{\label{betahistF} The distribution of $\beta$ values for the
sample, note that the value determined for the Galaxy with a
multi-component model is 1.5--2.0}
\end{figure}

The value of $\beta$ is remarkably uniform (the overall S.D. for the
whole sample is of the same order as the individual uncertainties,
derived from the above technique), and lower than the value of
1.5--2.0 obtained from multi-wavelength studies of our Galaxy (Masi et
al. 1995; Reach et al. 1995), and also of NGC 891 (Alton et
al. 1998b). The number of galaxies at $\beta \geq 1.5$ is equivalent
to those at $\beta \leq 1.1$, and in fact, the distribution of $\beta$
about the mean is well represented by a Gaussian, with those galaxies
at $\beta > 1.5$ being the tail of that Gaussian. The
distribution of $\beta$ values for the sample is shown in
Fig.~\ref{betahistF}. In fact, when the fluxes of the Milky Way and
NGC 891 are fitted in the same way as we have fitted the {\em IRAS\/}
galaxies (i.e. a single temperature model using only the 60,100 and
850$\mu$m points) the $\beta$ value we find is 0.7. It is only when
the full SED is used (which inlcudes fluxes between 200 and 800$\mu$m) 
that the colder component at $< 20$ K can be
identified, leading to the higher $\beta$ of 1.5--2. So rather than
implying that $\beta$ truly does have a low value, our results
probably indicate that there is dust in these galaxies at colder
temperatures than is indicated by a single-component fit. The
uniformity of the $\beta$ values in the sample could be indicative of
similar cold component properties in all of the {\em IRAS\/}
galaxies. 

\begin{figure}
 \vspace{6cm}
 \includegraphics{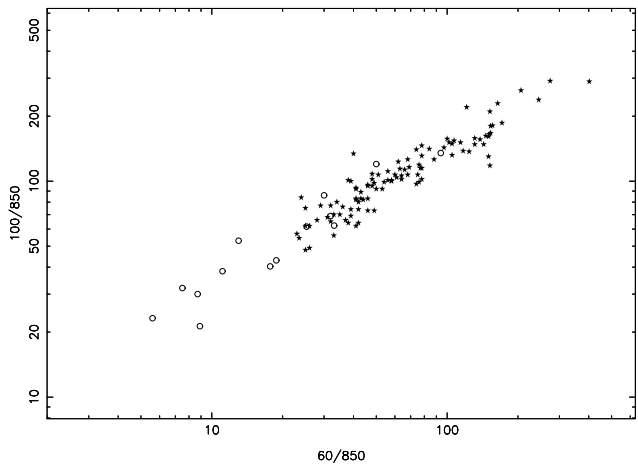}
 \caption{\label{colplotF} A plot of the 100/850 versus 60/850 colours
for the {\em IRAS\/} sample described in this paper (solid symbols)
plus the optically selected galaxies observed so far (open
symbols). The two samples clearly have different FIR-submm properties.}
\end{figure}

The lower $\beta$ of 0.7 determined in this way for the
Milky Way and NGC 891, which are optically selected, may be due to
them having larger fractions of colder dust than the {\em
IRAS\/} galaxies. We can partially test this using preliminary data 
from the optically selected sample. Figure~\ref{colplotF} shows a
colour-colour plot for the {\em IRAS\/} galaxies (stars) plus the galaxies
observed so far from the optical sample (circles). There is clear evidence that
the optically selected galaxies occupy a different area of the space,
implying that they are not merely the less luminous equivalent of {\em
IRAS\/} galaxies. We can make a more thorough investigation
when the optically selected sample is completed and we have 350 and
450$\mu$m data for the samples. 

\subsection{Dust masses}

Using the 850$\mu$m flux and the temperature derived in
the above way the dust mass is calculated from
 
\begin{equation}
M_{\rm{d}} = \frac{S_{850}\,D^2}{\kappa_d(\nu)\,B(\nu,\, T_{\rm{d}})}
\label{MdE}
\end{equation}

\parindent=0pt
where $\kappa_d(\nu)$ is the dust mass opacity
coefficient, $B(\nu,\, T_{\rm{d}})$ is the
value of the Planck function at 850$\mu$m for a temperature
$T_{\rm{d}}$, and $D$ for $\Omega_0 =$1 is given by

\begin{equation}
D = \frac{2c}{H_0}\left(1-\frac{1}{\sqrt{1+z}}\right) \label{DE}
\end{equation}
\parindent=1em

The dust masses are listed in Table~\ref{massT}, with values for
individual members of pairs in Table~\ref{pmassT}. We have assumed a
value of 0.077 m$^2$ kg$^{-1}$ for $\kappa_d(\nu)$, which is
intermediate between values for graphite and silicates as given by
Draine \& Lee (1984) and by Hughes et al. (1993), and also that
$\kappa_d(\nu)$ has the same value at 850$\mu$m for all our
galaxies. Even though the value of $\kappa_d(\nu)$ is notoriously
uncertain, the relative values of our dust masses will be correct as
long as the dust has similar properties in all galaxies. The
uncertainties in the relative dust masses then depend only on the
errors in $S_{850}$ and $T_{\rm{d}}$. The formal statistical
uncertainties in $T_{\rm{d}}$ are only $\sim$ a few K, but there is
the possibility that our assumption of a single temperature has biased
our estimates of $T_{\rm{d}}$ to higher values, and thus our mass
estimates to lower values. We can estimate the size of this effect in
the following way: A single temperature fit will produce a lower value
for $\beta$ than is actually true if more than one component is
present. In their {\em COBE\/}--FIRAS study of our Galaxy, Reach et
al. (1995) concluded that the highest observed value of $\beta$ in any
region was likely to be closest to the true value, with any excess
emission in other regions due to a colder component. If we adopt the
same approach and assume that all our galaxies have a true $\beta$ of
2 (our highest observed value is 1.9) and a uniform cold dust
temperature of 20 K, we can then fit a two-component temperature model
and calculate a new dust mass. We find that for the same mass
coefficient as before, the masses of dust estimated from this
two-component model are between 1.5 and 3 times higher than those
derived from a single temperature fit. The discrepancy is larger for
galaxies with lower $\beta$ and higher temperatures and some galaxies
which have steep SEDs and cold temperatures have little extra dust
when modelled in this way (e.g. NGC 5371, NGC 772). The dust masses
calculated assuming this form for the cold component are listed
alongside the single temperature masses in Table~\ref{massT}. They
will be referred to as `cold dust masses' in future.

\subsection{Gas masses}

Neutral hydrogen (H{\sc i}) fluxes were taken from the
literature\footnote{See notes to Table~\ref{massT}} and converted to
masses (in solar units) using

\begin{equation}
M_{\rm{HI}}=2.36\times 10^5\, D^2\, S_{\rm{HI}}     \label{HImE}
\end{equation}
\parindent=0pt

where $D^2$ (Eqn.~\ref{DE}) is in Mpc and $S_{\rm HI}$ is in Jy km s$^{-1}$\\
\parindent=1em

Molecular gas (H$_2$) masses were calculated using CO fluxes taken
from the literature$^3$ and scaled to telescope independent units (Jy
km s$^{-1}$). When an object appeared in more than one reference, an
average flux was used to determine the mass. Conversion factors used
for different telescopes are given in Table~\ref{convT}. Molecular
masses in M$_{\odot}$ are given by

\begin{equation}
M_{\rm{H_2}}=1.1\times 10^4\, D^2\, S_{\rm{CO}}   \label{H2mE}
\end{equation}
\parindent=0pt
(Kenny \& Young 1989), in which the parameters have the same units as in
Eqn.~\ref{HImE}. This assumes a CO to H$_2$ conversion factor of
$X=2.8\times10^{20}$ H$_2$ cm$^{-2}$/[K($T_R$)km s$^{-1}]$. Atomic and
molecular gas masses are also given in Table~\ref{massT}.

\subsection{Optical luminosities}

Blue magnitudes were taken from the Lyon-Meudon Extra-galactic Database
(LEDA) and converted to luminosities using $M_{B\odot}=5.48$. Blue
luminosities are given in Table~\ref{massT} and are corrected for
Galactic extinction but not for internal extinction or inclination
effects. In future discussions involving $L_{\rm{B}}$, mention will be
made of `corrected' blue luminosities. This will refer to the values
in Table~\ref{massT} after further correction for the internal effects
of dust in the galaxy, and for inclination, following the prescription
given in the 3rd Reference Catalogue of Bright Galaxies (RC3) (de
Vaucouleurs et al. 1991).

\subsection{Far Infra-Red Luminosities}

The FIR luminosity ($L_{\rm{fir}}$) is usually calculated using the 60
and 100$\mu$m {\em IRAS\/} fluxes, as described in the
Appendix of {\em Catalogued Galaxies and Quasars Observed in the IRAS
Survey\/} (Version 2, 1989):

\[ FIR = 1.26 \times 10^{-14}(2.58S_{60}+S_{100})\]

and

\[L_{\rm{fir}}=4\pi D^2 \times FIR \times C
\]

\input{table4}
\vspace{2mm}
where $D$ is defined in Eqn.~\ref{DE} and $C$ is a colour correction
factor which depends on the ratio of $S_{60}/S_{100}$ and the assumed
value of the emissivity index. The correction factor, designed to
account for emission outside the {\em IRAS\/} bands, varies between
1.3 and 2.4 and is explained in more detail by Helou et al. (1988).
Having submillimetre fluxes, we can use our derived temperatures and
$\beta$ to integrate the total flux under the SED directly out to
1000$\mu$m, hopefully leading to more accurate values for
$L_{\rm{fir}}$ since no general assumptions are being made. We
integrate from $40-1000\mu$m, since if we were integrating to shorter
wavelengths we should really include the {\em IRAS\/} 12 and 25$\mu$m
data points which would require a multi-component temperature model,
beyond the scope of what we are doing here. However, for our objects
the ratio of $S_{60}/S_{25}$ is greater than 2.4, meaning that the
contribution to the integral at $\lambda < 40\mu$m is not very
significant (if we integrate our SED with only the three data points
out to 1$\mu$m we find increases in $L_{\rm{fir}}$ of only a few per
cent). Our values for $L_{\rm{fir}}$ are similar to those calculated
using the standard {\em IRAS\/} formulation described above, although
slightly lower ($\sim 10$ per cent) than if the appropriate value of
$C$ were chosen for an emissivity index of one. This is probably
because calculating a temperature from the 60/100 flux ratio tends to
overestimate the temperature and therefore the correction factor
required. It must be noted that the true SED of these galaxies may not
be well represented by a single temperature model and the values of
$L_{\rm{fir}}$ would change accordingly. Our integrated values for
$L_{\rm{fir}}$ are listed in Table~\ref{massT}.

\input{table5}

\section{LUMINOSITY AND DUST MASS FUNCTIONS}

\input{table6}

\input{table7}

\begin{figure*}
 \vspace{6.5cm}
 \includegraphics{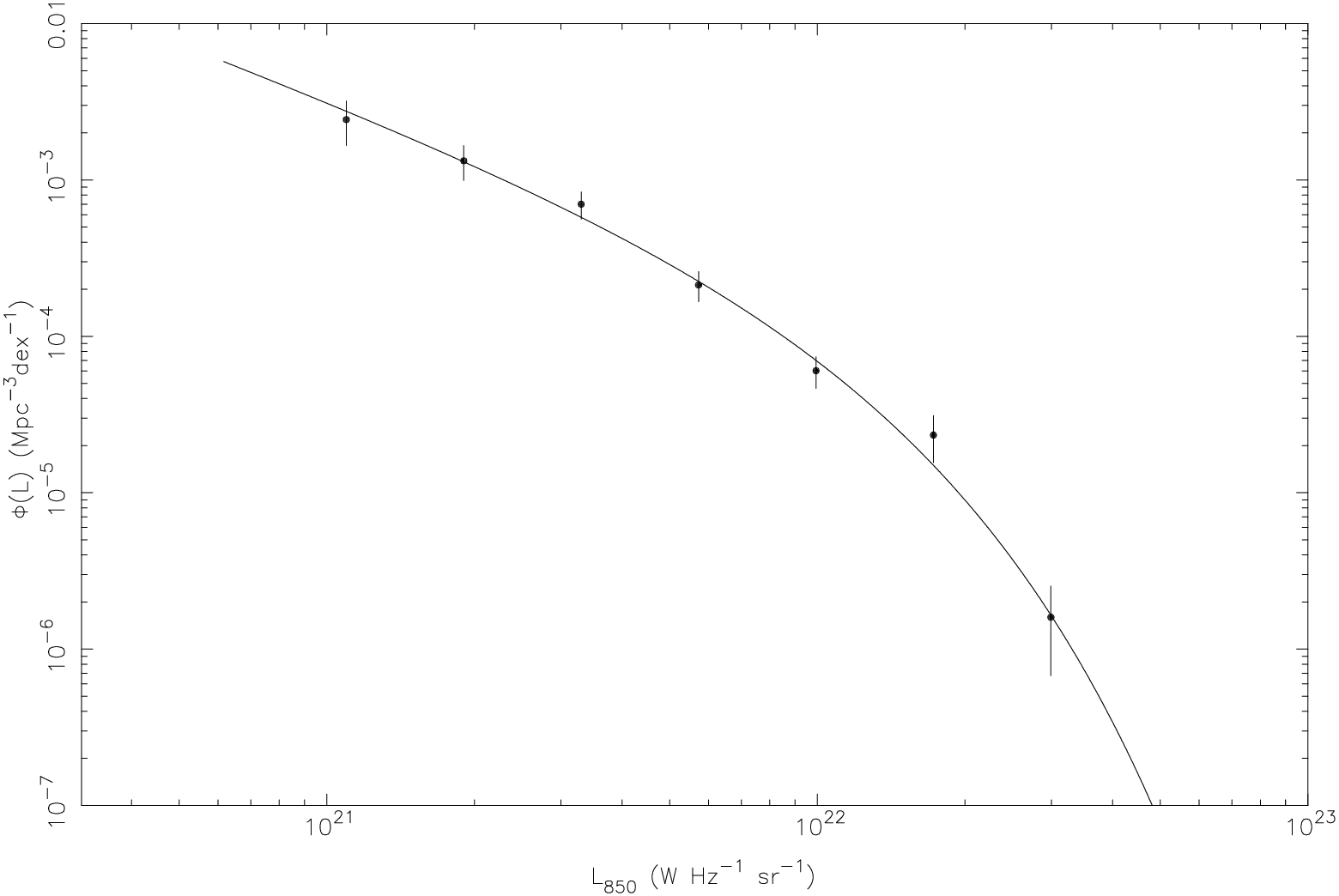}
 \includegraphics{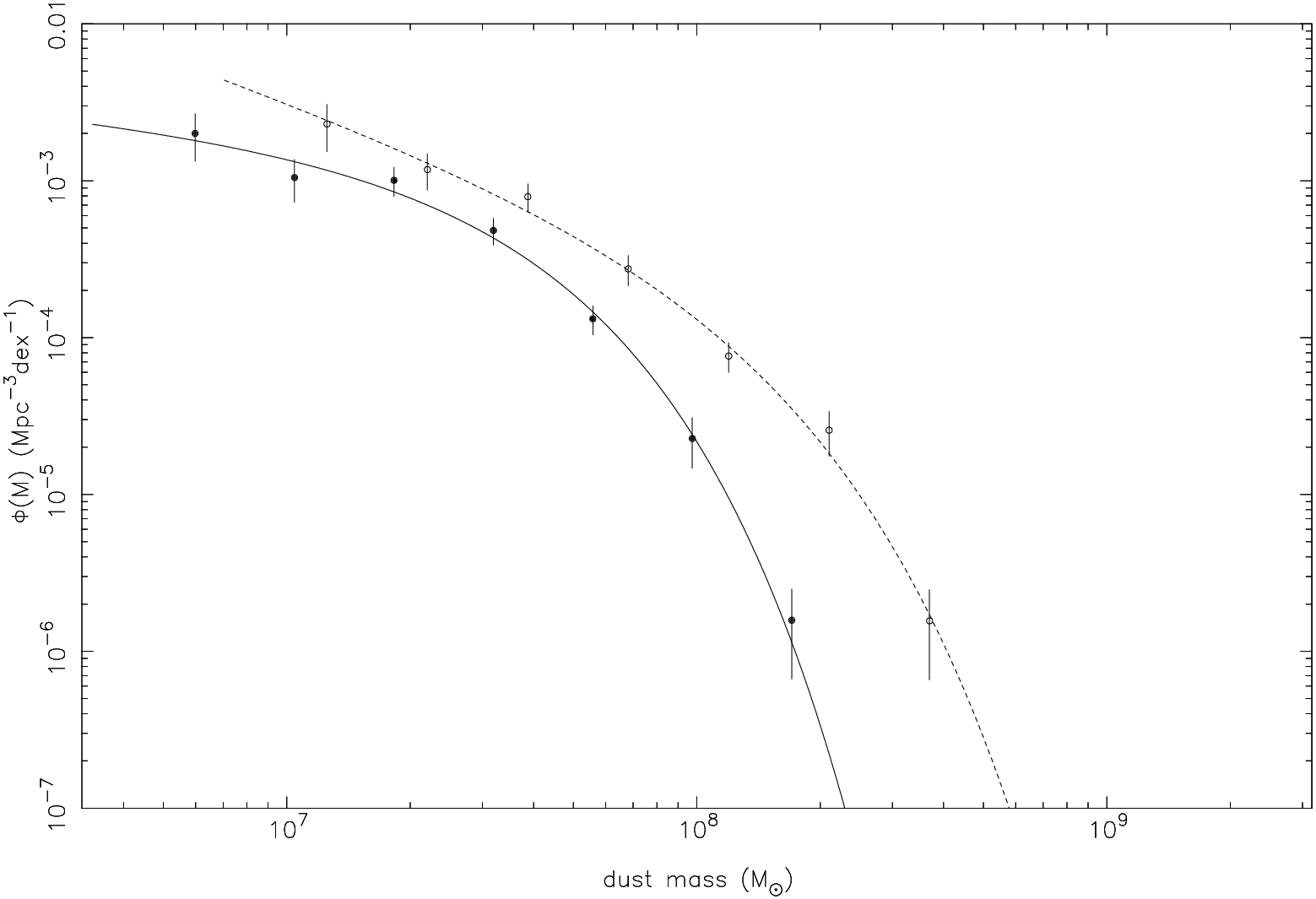}
\caption{\label{lf850F}a) Luminosity function at 850$\mu$m shown with
best-fitting Schechter function, $\alpha=-2.18$,
$L_{\ast}=8.3\times10^{21}$ W Hz$^{-1}$ sr$^{-1}$. b) Dust mass
function. Solid line:-- dust masses calculated from the submillimetre
fluxes using the best-fitting $\beta$ and $T_{\rm{d}}$. Schechter
parameters shown are  $\alpha=-1.23$,  $M_{\ast}=2.5\times10^7$
M$_{\odot}$. Dashed line:-- `cold dust mass' function using
$T_{\rm{cold}}=20\,\rm{K}$ and $\beta=2$. Schechter parameters are
$\alpha=-1.91$ and $M_{\ast}^{cold}=8.5\times10^{7}$ M$_{\odot}$.}
\end{figure*}

The luminosity function (LF) is given by

\begin{equation}
\Phi(L)\Delta L = \sum_i \frac{1}{V_i}     \label{lfE}
\end{equation}

where $\Phi(L)\Delta L$ is the number density of sources (Mpc$^{-3}$)
in the luminosity range $L$ to $L+\Delta L$, the sum is over all the
sources in the sample in this luminosity range, and $V_i$ is the
accessible volume of the {\em i}th source in the original sample (Avni
\& Bahcall 1980). For this sample, the accessible volume is the
maximum volume in which the object could be seen and still be in the
{\em IRAS} Bright Galaxy Sample and so to calculate this, the 60$\mu$m
luminosity of each galaxy is used. In Equation~\ref{lfE}, however, the
luminosity is the luminosity at 850$\mu$m, the wavelength of our
survey. The volume from $cz=0$ to $cz=$ 1900 km s$^{-1}$ is not included
in the calculation of $V_i$ as galaxies with velocities less than this
were excluded from our sample. $\Phi(L)$ has been normalised to
dex$^{-1}$ by dividing by $\Delta L$. The dust mass function is
estimated in the same way as the luminosity function but substituting
dust mass for luminosity in Eqn.~\ref{lfE}. This was done for both
values of the dust mass; using a single temperature, and assuming a colder
component at $T_{\rm{d}}=20$ K. The luminosity function at
850$\mu$m and both of the dust mass functions are shown in
Fig.~\ref{lf850F}, and tabular forms for the functions are given in
Table~\ref{lfT}.

\parindent=1em

\begin{figure}
 \vspace{8cm}
 \includegraphics{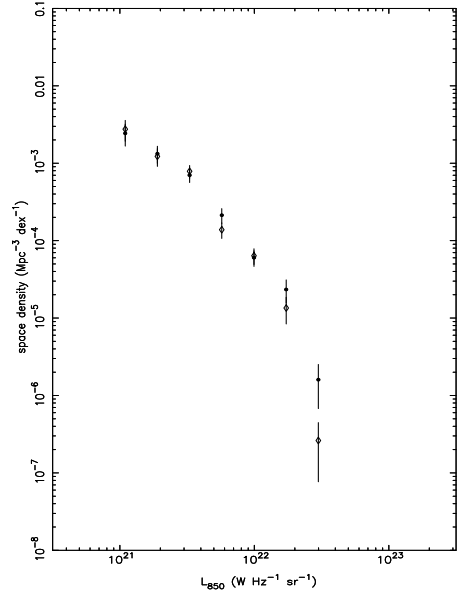}
\caption{\label{lfpF} Luminosity function at 850$\mu$m for combined
pair fluxes (Table~\ref{fluxT}) -- solid symbols, compared to the
luminosity function when the fluxes for the resolved pairs are
separated (Table~\ref{pfluxT}) -- open symbols.}
\end{figure}

The BGS contains many close pairs which are resolved at 850$\mu$m but
not at the {\em IRAS\/} wavelengths. A few do have HIRES fluxes
(Surace et al. 1993) and, when the resolved 60$\mu$m flux is still
above the 5.24 Jy limit of the sample, can be treated as separate
objects for the purposes of calculating the luminosity and dust mass
functions. The effect on the luminosity function of separating sources
in this way would be to steepen it, as one luminous source with a
large accessible volume becomes two less luminous sources with smaller
volumes. We attempted to quantify this in the following way. For
galaxies without HIRES fluxes, we estimated 60$\mu$m fluxes for the
individual galaxies in each pair using the FIR-radio correlation
(Helou et al. 1985). We removed any galaxy whose flux fell below the
flux limit of the sample and re-calculated the luminosity
function. Figure~\ref{lfpF} shows this luminosity function compared
with our original one (calculated using the sum of the 850$\mu$m
emission from a pair of galaxies unresolved at 60$\mu$m). There is
only a significant difference in the highest luminosity bin, which
suffers from noise anyway due to the small number of objects. In
future, we will refer only to the luminosity function constructed
using the combined pair fluxes (Fig.~\ref{lf850F}a), since this can be
most readily compared with the 60$\mu$m LF.

The luminosity function estimator (Eqn.~\ref{lfE}) is unbiased
providing there is no population of submillimetre emitting galaxies
which have a high space density, yet are completely absent from the
original {\em IRAS} sample. The only conceivable type of galaxy to
which this could apply would be a hypothetical `cold' population with
$T_{\rm{d}} < 25$ K. To investigate this possibility, let us assume
that there are two populations of galaxies with equal space
densities and 850$\mu$m luminosities, one with dust temperatures
at 20 K and the other at 35 K. The relative numbers in an 850$\mu$m
flux-limited survey would be given by\\

\[ \frac{N_{A}}{N_{B}}\, = \:\frac{\Phi_{A}}{\Phi_{B}}\times
\frac{V_{A}}{V_{B}}\, \approx \:
\frac{\Phi_{A}}{\Phi_{B}}\,\frac{(L_{850,A})^{3/2}}{(L_{850,B})^{3/2}}
\]
\parindent=0pt
where A and B refer to the two populations, $\Phi$ is the space
density, $V$ is the `accessible volume' in which one of the galaxies
would have been detected by the survey, and $L_{850}$ is the
luminosity at 850$\mu$m. Given the assumptions we have made, $N_A /
N_B =1$ and therefore equal numbers of the two populations should be
found by the survey.

\parindent=1em

Now let us consider the relative numbers that would be found by a
60$\mu$m flux-limited survey (the BGS). Replacing the 850$\mu$m
luminosity in the above equation with 60$\mu$m luminosity, we now
predict that the ratio $N_{35K} / N_{20K}$ will be $\sim
724$. Clearly, in a sample of $\sim100$ {\em IRAS\/} galaxies, finding
a member of this hypothetical cold population would be unlikely and so
our attempt to construct the 850$\mu$m LF from our
sample would be an underestimate. Making a similar calculation for galaxies
at $T_{\rm{d}}=25$ K and 35 K we now expect to find
$\sim$ 32 times as many warm galaxies as cold ones, so in our sample
we would expect to see around 3 objects at $T_{\rm{d}}=25$ K. There
are actually two galaxies with $T_{\rm{d}} \sim25$ K in our sample,
which is consistent with the idea that we could be missing a significant
cold population.

Is there truly a possibility that there is a missing population of
galaxies? An important point to realise is that our dust temperatures
are merely those corresponding to a best-fitting single-component
model and are not {\em actual\/} dust temperatures. One way to assess
the possibility that we are missing galaxies is to carry out precisely
the same fitting procedure for optically-selected galaxies. For
example, NGC 891 has been studied at many submillimetre wavelengths
(Alton et al. 1998b), but if we throw away all the measurements except
those at 60, 100 and 850$\mu$m, and carry out our fitting procedure,
we obtain $T_{\rm{d}}=34$ K and $\beta =0.7$. While this temperature
is similar to the average temperature for our sample, indicating that
there is a warm component in NGC 891, the value of 0.7 for $\beta$ is
significantly lower than for most of our galaxies suggesting a large
amount of colder dust, and when the full submillimetre data-set was
used the bulk of the dust in NGC 891 was found to be at 15 K. Consider now
our own Galaxy : if the {\em IRAS\/} FIR and ARGO submillimetre data
(Sodroski et al. 1989; Masi et al. 1995) for the Milky Way is combined
and fitted in our usual way we find a temperature of 28 K with a
$\beta$ of 0.7, so again the SED of the Milky Way would not
necessarily have excluded it from our sample. The question of whether
we are missing cold galaxies remains open (primarily how many cold
galaxies there are to miss). To address this properly, and to
constrain the luminosity function better at low luminosities we will
need to complete the survey of an optically-selected sample, which
will not be biased by temperature selection as is the
BGS.

\begin{figure}
 \vspace{8.8cm}
 \includegraphics{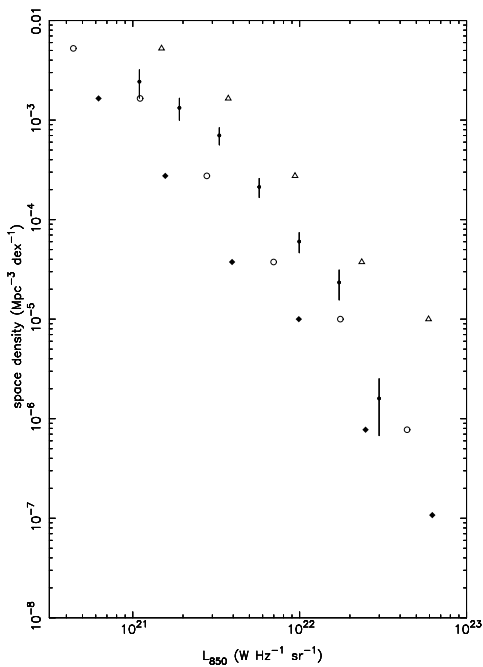}
\caption{\label{602850F} The measured 850$\mu$m luminosity
function (solid symbols with error bars) along with extrapolations of the
60$\mu$m LF from Soifer et al. (1987), using the following fixed
parameters:- $\beta=2,\,T_{\rm{d}}=24$ K -- open triangles; $\beta=1.5,
\, T_{\rm{d}}=38$ K -- solid diamonds; $\beta=1, \,T_{\rm{d}}=45$ K -- open
circles.}
\end{figure}

\begin{figure}
 \vspace{8.8cm}
 \includegraphics{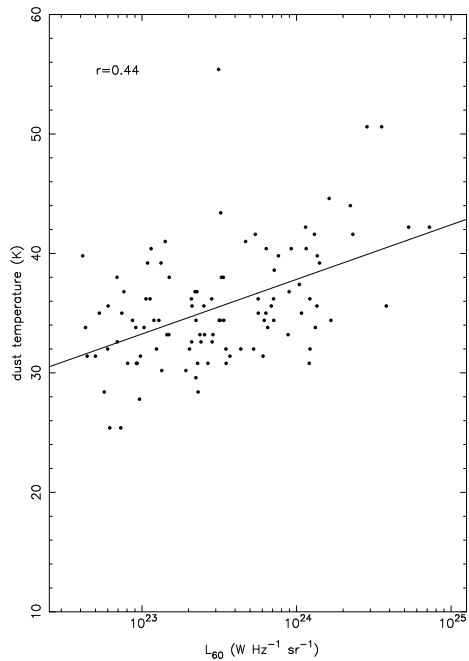}
\caption{\label{60TF} Dust temperature $T_{\rm{d}}$ versus 60$\mu$m
luminosity, showing that the most luminous galaxies have the hottest
temperatures, the abrupt cutoff at the low luminosity end is due to
the lower velocity limit in the selection of the sample at 60$\mu$m.}
\end{figure}

\begin{figure}
 \vspace{8.8cm}
 \includegraphics{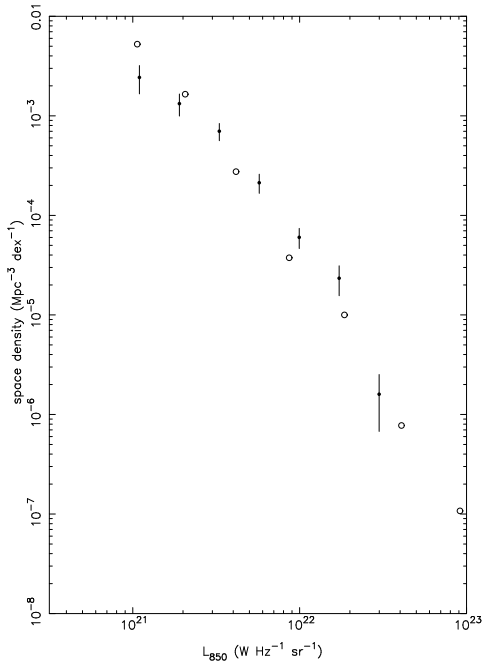}
\caption{\label{602850bF} 850$\mu$m LF (solid symbols)
along with the extrapolation of the Soifer et al. (1987) 60$\mu$m LF
using variable $\beta$ and $T_{\rm{d}}$ given by the relationships in
Table~\ref{fitsT}. This produces a better match to the measured
850$\mu$m LF than extrapolations made using single values for
$T_{\rm{d}}$ and $\beta$ (see Fig.~\ref{602850F}).}
\end{figure}

The 60$\mu$m luminosity function of the BGS from Soifer et al. (1987),
can be compared to the 850$\mu$m LF by making an extrapolation based
on an assumption about a galaxy's SED from the far-infrared to the
submillimetre. In fact, all of the 60$\mu$m LFs using {\em IRAS} data
are consistent with each other (with the exception the one from Saunders et
al. (1990), see Lawrence et al. 1999) and so it does not really matter
which one we use. We have used the Soifer et al. one as it represents
the same sample of galaxies which we observed. In the past this
extrapolation has been done by assuming a single temperature and
$\beta$ for all galaxies, which for 60 and 850$\mu$m gives the
following

\begin{equation}
L_{850}=L_{60} \times \left(\frac{\nu_{850}}{\nu_{60}}\right)^{3+\beta}
\times
\left(\frac{e^{240.2/T_{\rm{d}}}-1}{e^{16.8/T_{\rm{d}}}-1}\right)
\label{602850E}
\end{equation}
\parindent=0pt
Extrapolations of the 60$\mu$m LF using a range of plausible values of
$\beta$ and $T_{\rm{d}}$ are shown over the measured 850$\mu$m LF in
Fig~\ref{602850F}. As one would expect, an 850$\mu$m luminosity
function obtained by extrapolating in wavelength from a 60$\mu$m LF
is highly dependent on the values assumed for $T_{\rm{d}}$ and
$\beta$. In practice, luminosity functions obtained in this way have
had much lower amplitudes than the one we actually measure,
essentially because dust temperatures deduced from {\em IRAS\/} data
alone are always higher than the temperatures we estimate here. This
can have significant consequences when trying to model the evolution
required to fit the observed submillimetre number counts from deep
surveys and the submillimetre background and the implications will be
discussed in a future paper (Eales et al. in prep). The other
difference in the luminosity functions is the slope: the luminosity
function we measure is steeper than the ones extrapolated from the
60$\mu$m LF. The difference in slope between the two LFs is due to the
correlation of dust temperature with 60$\mu$m luminosity
(Fig.~\ref{60TF}). As the temperature changes with 60$\mu$m
luminosity, extrapolating every $L_{60}$ to 850$\mu$m with an average
$T_{\rm{d}}$ will underestimate the 850$\mu$m luminosity at lower
$L_{60}$ and overestimate it at high $L_{60}$. Using the fitted
dependences of $L_{60}$ on $T_{\rm{d}}$ and also $\beta$ on
$T_{\rm{d}}$ (see Table~\ref{fitsT}) we again extrapolate the 60$\mu$m
LF and this time the agreement with the 850$\mu$m LF is much better
(Fig.~\ref{602850bF}).

The 850$\mu$m luminosity and dust mass functions are well fitted by
Schechter functions of the form (Press \& Schechter 1974; Schechter 1975)

\[ 
\Phi(L)dL =\phi(L)\left(\frac{L}{L_{\ast}}\right)^{\alpha}e^{-(L/L{\ast})}\,dL/L_{\ast} 
\]
 
\begin{figure*}
 \vspace{5cm}
 \includegraphics{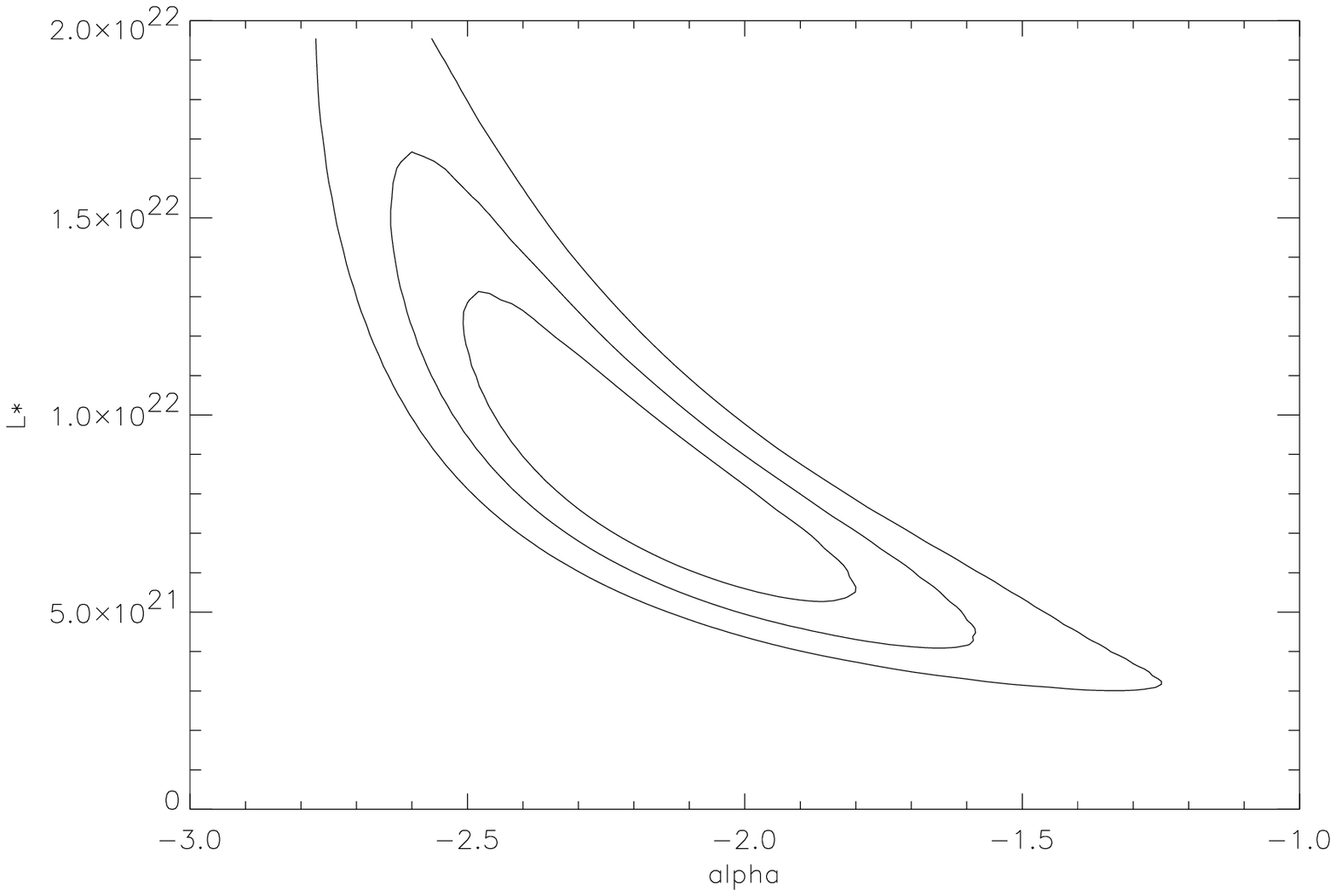}
 \includegraphics{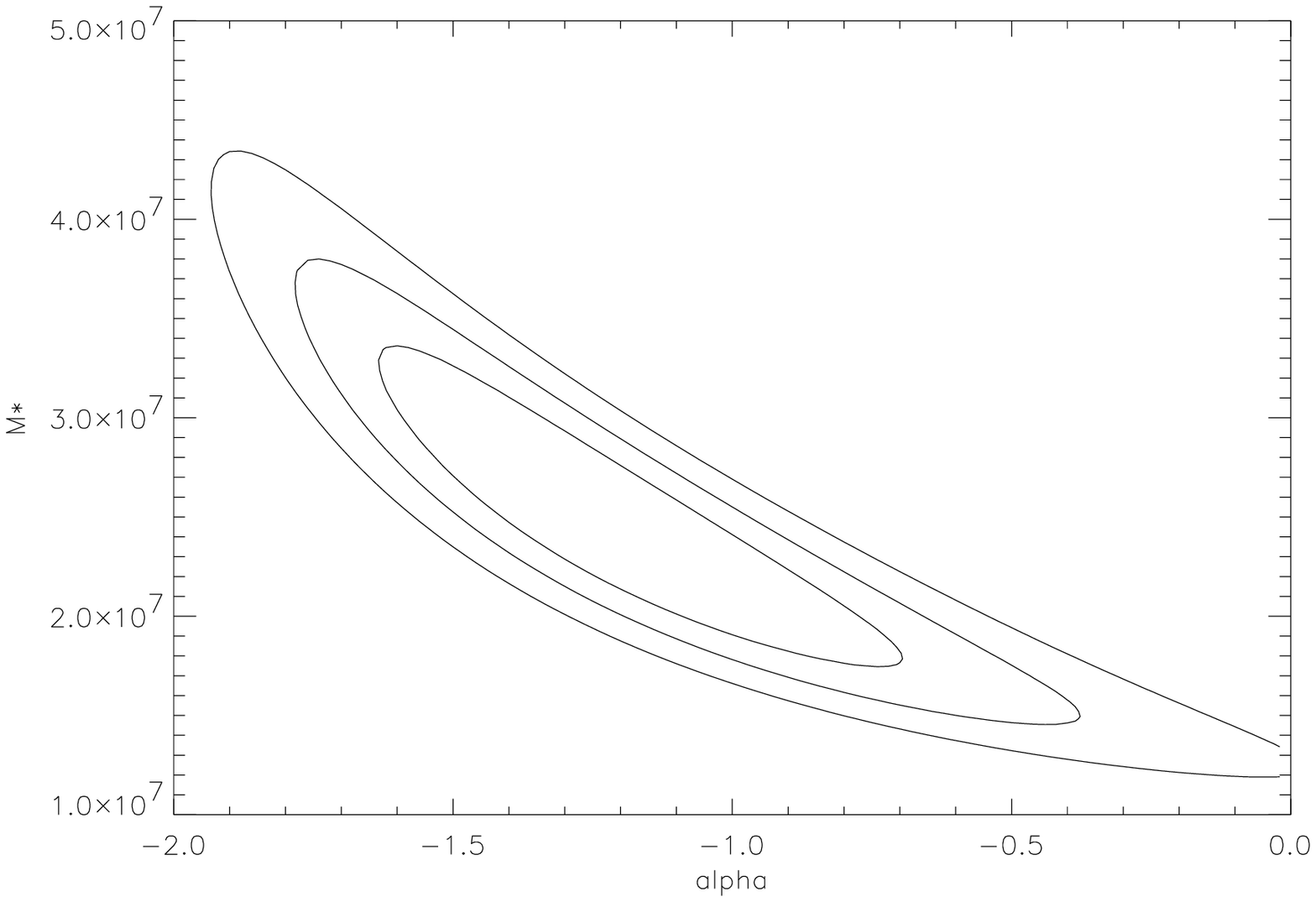}
 \includegraphics{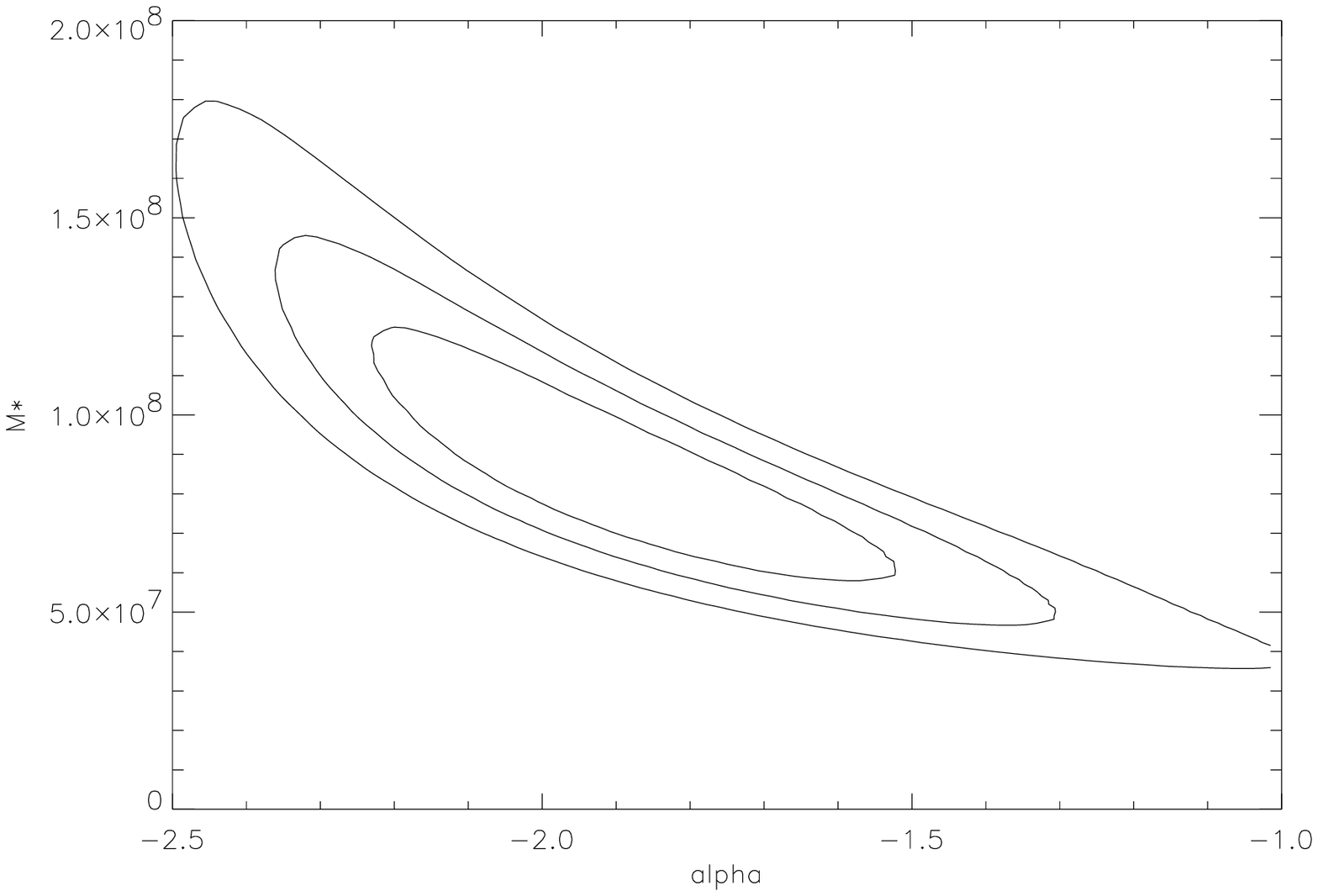}
\caption{\label{chicontF}a) Joint confidence $\chi^2$contours for the
850$\mu$m LF parameters $\alpha$ and $L_{\ast}$. Contours are at the
68, 90 and 95\% confidence level, d.o.f = 2. b) Joint confidence
contours for the single temperature dust mass function parameters
$\alpha$ and $M_{\ast}$. Same contour levels as a). c) Joint
confidence contours for the `cold' dust mass function, same parameters
and levels as b).}
\end{figure*}

The best fitting parameters for the 850$\mu$m LF and the dust mass
functions along with the reduced chi-squared values ($\chi^{2}_{\nu}$) 
for the fits, are given in Table~\ref{lfT}. The $\chi^2$ contours
showing the joint confidence intervals on $\alpha$ and $L_{\ast}$ for
all the functions are shown in Figs.~\ref{chicontF}(a--c).

There are several points to note here. Firstly, the 60$\mu$m LF cannot
be fitted by a simple Schechter function (Lawrence et al. 1986; Rieke
\& Lebofsky 1986) as the high luminosity end does not fall off steeply 
enough. The fact that the 850$\mu$m LF and the dust mass
function can, suggests that the Schechter function is surprisingly
universal and that in the case of the 60$\mu$m LF, the Schechter
function exponential fall-off is concealed by the sensitivity of the
60$\mu$m emission to dust temperature. Secondly, the slope of the
850$\mu$m LF is steeper than $-2$ ($\alpha=-2.18$ ) at lower submillimetre
luminosities. This is significant because if this slope continued to
zero luminosity, the submillimetre sky would be infinitely bright (a
submillimetre Olbers Paradox). Of course, this means that the slope
must flatten out at lower luminosities not yet probed by our
survey. The shape of the dust mass function is strongly affected by
the temperature distribution assumed, i.e. all of the dust at a single
temperature, or most of the dust at some colder temperature with a
small warmer component being responsible for the 60$\mu$m flux. The
shape of the `cold dust mass' function bears more resemblance to the
850$\mu$m LF simply because of our assumption of a common universal
temperature for the cold component. We do believe however, that the
true distribution of the dust masses lies somewhere between the two;
the temperature of a cold component may not be the same in all
galaxies but will probably vary less than the temperature of the warm
components and this degree of variation along with the relative
amounts of cold and warm dust will determine how similar the dust mass
function is in shape to the 850$\mu$m LF.


%% file: table4.tex

\onecolumn
\begin{longtable}{lcccccccccc}
\caption{\label{massT} Luminosities and masses}\\
\hline
\\[-2.5ex]
\multicolumn{1}{c}{Name}&\multicolumn{1}{c}{log $L_{60}$}&\multicolumn{1}{c}{log
$L_{850}$}&\multicolumn{1}{c}{log $L_{\rm{fir}}$}&\multicolumn{1}{c}{log $M_{\rm{d}}$}&\multicolumn{1}{c}{log
$M_{\rm{d}}^{cold}$}&\multicolumn{1}{c}{log
$M_{\rm{HI}}$}&\multicolumn{1}{c}{log
$M_{\rm{H_2}}$}&\multicolumn{1}{c}{log
$L_{\rm{B}}$}&\multicolumn{1}{c}{$G_{\rm{d}}$}&\multicolumn{1}{c}{$L_{\rm{fir}}/M_{\rm{H_2}}$}\\
\\[-2ex]
\multicolumn{1}{c}{}&\multicolumn{1}{c}{(W
Hz$^{-1}$}&\multicolumn{1}{c}{(W Hz$^{-1}$}&\multicolumn{1}{c}{(L$_{\odot}$)}&\multicolumn{1}{c}{(M$_{\odot}$)}&\multicolumn{1}{c}{(M$_{\odot}$)}&\multicolumn{1}{c}{(M$_{\odot}$)}&\multicolumn{1}{c}{(M$_{\odot}$)}&\multicolumn{1}{c}{(L$_{\odot}$)}&\multicolumn{1}{c}{}&\multicolumn{1}{c}{(L$_{\odot}$/M$_{\odot}$)}\\
\multicolumn{1}{c}{}&\multicolumn{1}{c}{sr$^{-1}$)}&\multicolumn{1}{c}{sr$^{-1}$)}&\multicolumn{7}{c}{}\\
\\[-2ex]
\multicolumn{1}{c}{}&\multicolumn{1}{c}{(1)}&\multicolumn{1}{c}{(2)}&\multicolumn{1}{c}{(3)}&\multicolumn{1}{c}{(4)}&\multicolumn{1}{c}{(5)}&\multicolumn{1}{c}{(6)}&\multicolumn{1}{c}{(7)}&\multicolumn{1}{c}{(8)}&\multicolumn{1}{c}{(9)}&\multicolumn{1}{c}{(10)}\\
\\[-2.5ex]
\hline
\\[-2.5ex]
\endfirsthead
\caption{ ... continued}\\
\hline
\\[-2.5ex]
\multicolumn{1}{c}{Name}&\multicolumn{1}{c}{log $L_{60}$}&\multicolumn{1}{c}{log
L$_{850}$}&\multicolumn{1}{c}{log $L_{\rm{fir}}$}&\multicolumn{1}{c}{log $M_{\rm{d}}$}&\multicolumn{1}{c}{log
$M_{\rm{d}}^{cold}$}&\multicolumn{1}{c}{log
$M_{\rm{HI}}$}&\multicolumn{1}{c}{log $M_{\rm{H_2}}$}&\multicolumn{1}{c}{log
$L_{\rm{B}}$}&\multicolumn{1}{c}{$G_{\rm{d}}$}&\multicolumn{1}{c}{$L_{\rm{fir}}/M_{\rm{H_2}}$}\\
\\[-2ex]
\multicolumn{1}{c}{}&\multicolumn{1}{c}{(W
Hz$^{-1}$}&\multicolumn{1}{c}{(W
Hz$^{-1}$}&\multicolumn{1}{c}{(L$_{\odot}$)}&\multicolumn{1}{c}{(M$_{\odot}$)}&\multicolumn{1}{c}{(M$_{\odot}$)}&\multicolumn{1}{c}{(M$_{\odot}$)}&\multicolumn{1}{c}{(M$_{\odot}$)}&\multicolumn{1}{c}{(L$_{\odot}$)}&\multicolumn{1}{c}{}&\multicolumn{1}{c}{(L$_{\odot}$/M$_{\odot}$)}\\
\multicolumn{1}{c}{}&\multicolumn{1}{c}{sr$^{-1}$)}&\multicolumn{1}{c}{sr$^{-1}$)}&\multicolumn{7}{c}{}\\
\\[-2ex]
\multicolumn{1}{c}{}&\multicolumn{1}{c}{(1)}&\multicolumn{1}{c}{(2)}&\multicolumn{1}{c}{(3)}&\multicolumn{1}{c}{(4)}&\multicolumn{1}{c}{(5)}&\multicolumn{1}{c}{(6)}&\multicolumn{1}{c}{(7)}&\multicolumn{1}{c}{(8)}&\multicolumn{1}{c}{(9)}&\multicolumn{1}{c}{(10)}\\
\\[-2.5ex]
\hline
\\[-2.5ex]
\endhead
\\[-2.5ex]
\hline
\endfoot
\\[-2.5ex]
\hline
\endlastfoot

NGC 23 & 23.51 & 21.68 & 10.82 & 7.49 & 7.79 & 9.88 & 9.95 & 10.73 & 746 & 7.5\\
UGC 556 & 23.31 & 21.43 & 10.86 & 7.28 & 7.57 & 9.64 & 9.61 & 9.72 & 444 & 10.9\\
NGC 470 & 23.32 & 21.30 & 10.16 & 7.05 & 7.43 & 9.61 & 9.29 & 10.30 & 536 & 7.4\\
MCG+02-04-025 & 24.21 & 21.71 & 11.38 & 7.39 & 7.76 & 9.54 &...& 10.40 & ... & ...\\
UGC 903 & 22.94 & 21.27 & 10.27 & 7.08 & 7.38 & 9.52 &...& 9.44 & ... & ...\\
NGC 520 & 23.45 & 21.45 & 10.73 & 7.22 & 7.53 & 9.77 & 9.88 & 10.35 & 809 & 7.0\\
III ZW 035 & 24.15 & 21.90 & 11.36 & 7.64 & 7.96 & 9.36 & 9.90 & 9.97 & 238 & 28.6\\
NGC 695 & 24.12 & 22.29 & 11.42 & 8.11 & 8.41 & 10.33 & 10.55 & 10.86 & 443 & 7.8\\
NGC 697 & 22.98 & 21.55 & 10.49 & 7.47 & 7.66 & 10.37 & ...& 10.43 & ... & ...\\
UGC 1351 & 23.35 & 21.67 & 10.68 & 7.48 & 7.79 & 9.79 & ...& 10.24 & ... & ...\\
UGC 1451 & 23.46 & 21.62 & 10.78 & 7.44 & 7.71 & 9.63 & ...& 10.11 & ... & ...\\
NGC 772 & 22.86 & 21.46 & 10.46 & 7.44 & 7.56 & 10.36 & 10.15 & 10.81 & 1344 & 2.0\\
NGC 877 & 23.50 & 21.92 & 10.85 & 7.72 & 8.04 & 10.31 & 10.00 & 10.67 & 573 & 7.2\\
NGC 958 & 23.54 & 22.14 & 10.97 & 8.01 & 8.27 & 10.59 & 10.21 & 10.75 & 544 & 5.8\\
NGC 992 & 23.51 & 21.61 & 10.78 & 7.36 & 7.72 & 10.07 & 9.83 & 9.58 & 809 & 8.8\\
UGC 2238 & 23.78 & 21.83 & 11.11 & 7.69 & 7.90 & 10.02 & 10.22 & 10.20 & 550 & 7.9\\
IR 0243+21 & 23.67 & 21.48 & 10.88 & 7.20 & 7.58 & 9.13 & ...& ...& ... & ...\\
NGC 1134 & 23.32 & 21.72 & 10.65 & 7.50 & 7.85 & 10.28 & ...& 10.34 & ... & ...\\
UGC 2369 & 24.09 & 21.98 & 11.33 & 7.77 & 8.07 &... & ...& 10.49 & ... & ...\\
UGC 2403 & 23.36 & 21.49 & 10.64 & 7.26 & 7.60 & ...& ...& 9.82 & ... & ...\\
NGC 1222 & 23.12 & 20.92 & 10.36 & 6.66 & 6.98 & 9.38 & ...& 9.98 & ... & ...\\
IR 0335+15 & 24.06 & 21.87 & 11.25 & 7.57 & 7.98 &...&10.44 & 9.95 & ... & 6.4\\
UGC 2982 & 23.64 & 21.90 & 10.99 & 7.74 & 8.00 & 10.07 & 10.23 & 9.98 & 520 & 5.8\\
NGC 1614 & 24.12 & 21.90 & 11.33 & 7.61 & 8.00 & 9.54 & 10.09 & 10.34 & 389 & 17.2\\
NGC 1667 & 23.36 & 21.73 & 10.81 & 7.65 & 7.82 & 9.75 & 9.98 & 10.50 & 342 & 6.7\\
NGC 2623 & 24.13 & 21.66 & 11.33 & 7.39 & 7.67 & 9.25 & 9.97 & 10.32 & 456 & 22.7\\
IR 0857+39 & 24.55 & 21.85 & 11.68 & 7.46 & 8.00 &...& 9.69 & 11.12 & ... & 98.0\\
NGC 2782 & 23.04 & 21.41 & 10.34 & 7.14 & 7.54 & 9.62 & 9.36 & 10.45 & 465 & 9.5\\
NGC 2785 & 23.08 & 21.39 & 10.42 & 7.20 & 7.51 & 8.99 &...& 9.42 & ... & ...\\
UGC 4881 & 24.22 & 22.12 & 11.50 & 7.94 & 8.21 & 10.25 & 10.65 & 10.68 & 723 & 6.6\\
NGC 2856 & 22.87 & 21.01 & 10.18 & 6.80 & 7.11 & 9.15 & ... & ... & ...\\
MCG+08-18-012 & 23.84 & 21.59 & 11.07 & 7.38 & 7.65 & ...& ...& 10.21 & ...&...\\
NGC 2966 & 22.62 & 20.97 & 9.91 & 6.70 & 7.10 & 9.14 & ...& 9.63 & ... & ...\\
NGC 2990 & 22.96 & 21.23 & 10.30 & 7.05 & 7.35 & 9.69 & ... & 10.05 & ... & ...\\
IC 563/4 & 23.54 & 22.11 & 10.95 & 7.96 & 8.25 & 10.10 & ... & 10.53 & ...&...\\
UGC 5376 & 22.63 & 21.01 & 10.00 & 6.83 & 7.13 & 9.41 & ... & 9.54 & ...&...\\
NGC 3094 & 23.06 & 21.16 & 10.31 & 6.88 & 7.27 & 9.35 & ...& 9.99 & ...&...\\
NGC 3110 & 23.72 & 21.88 & 11.06 & 7.73 & 7.98 & ...& 10.34 & 10.52 & ...& 5.2\\
IR 1017+08 & 24.35 & 22.03 & 11.52 & 7.71 & 8.16 & ...& 10.38 & ... & ...& 13.8\\
NGC 3221 & 23.35 & 21.84 & 10.79 & 7.73 & 7.96 & 10.33 & 9.99 & 10.14 & 581 & 6.3\\
NGC 3367 & 22.99 & 21.30 & 10.38 & 7.16 & 7.40 & 9.77 &... & 10.68 & ... & ...\\
IR 1056+24 & 24.58 & 22.16 & 11.78 & 7.96 & 8.18 &... & 10.36 & 10.92 & ...& 26.2\\
ARP 148 & 24.13 & 22.17 & 11.39 & 7.96 & 8.32 & ...& 10.21 & ... & ... & 15.4\\
NGC 3583 & 22.96 & 21.14 & 10.22 & 7.06 & 7.23 & 9.68 & 9.49 & 10.49 & 694 & 5.3\\
MCG+00-20-023 & 23.75 & 21.88 & 11.04 & 7.68 & 7.99 &...& ...& 10.38 & ... &...\\
UGC 6436 & 24.03 & 22.23 & 11.32 & 8.02 & 8.36 & 9.88 & 10.30 & 10.45 & 260 & 10.5\\
NGC 3994/5 & 23.16 & 21.58 & 10.53 & 7.41 & 7.70 & 10.25 & ... & 10.65 & ...& ...\\
NGC 4045 & 22.64 & 20.97 & 10.04 & 6.82 & 7.07 & 9.13 & ...& 10.02 & ... &...\\
IR 1211+03 & 24.86 & 22.48 & 12.00 & 8.19 & 8.60 &...& 10.64 & 10.47 & ...& 22.9\\
NGC 4273 & 23.01 & 21.48 & 10.39 & 7.30 & 7.61 & 9.64 & 9.63 & 10.37 & 435 & 5.8\\
IR 1222-06 & 23.80 & 21.86 & 11.04 & 7.58 & 7.98 & ...& ...& 9.84 & ... & ...\\
NGC 4418 & 23.50 & 21.30 & 10.68 & 6.86 & 7.41 & 8.75 & 9.12 & 9.68 & 261& 36.4\\
NGC 4433 & 23.34 & 21.51 & 10.64 & 7.28 & 7.62 & 9.86 & 9.63 & 10.20 & 600 & 10.2\\
NGC 4793 & 23.12 & 21.42 & 10.54 & 7.30 & 7.51 & 9.79 & 9.67 & 10.41 & 550 & 7.4\\
NGC 4922 & 23.73 & 21.62 & 10.95 & 7.32 & 7.73 & 9.15 & ...& 10.67 & ... & ...\\
NGC 5020 & 23.03 & 21.58 & 10.38 & 7.36 & 7.71 & 10.07 & ...& 10.44 & ...& ...\\
IC 860 & 23.51 & 21.33 & 10.72 & 7.01 & 7.43 &... & 8.91 & 9.68 & ...& 64.1\\
UGC 8387 & 24.08 & 21.93 & 11.39 & 7.81 & 7.96 & ...& 10.17 & 10.35 & ...& 16.5\\
NGC 5104 & 23.57 & 21.65 & 10.90 & 7.51 & 7.74 & 9.83 & ...& 10.34 & ...& ...\\
NGC 5256 & 23.95 & 21.94 & 11.20 & 7.72 & 8.04 & ...& 10.36 & 10.87 & ...& 6.9\\
NGC 5257/8 & 23.94 & 22.31 & 11.24 & 8.14 & 8.44 & 10.55 & ...& 11.04 & ...&...\\
UGC 8739 & 23.43 & 21.88 & 10.84 & 7.75 & 8.00 & 10.12 & ... & 10.47 & ...&...\\
NGC 5371 & 22.79 & 21.16 & 10.34 & 7.14 & 7.18 & 10.05 & ...& 10.80 & ...&...\\
NGC 5394/5 & 23.28 & 21.72 & 10.71 & 7.60 & 7.84 & 10.35 & ... & 10.86 & ...& ...\\
NGC 5433 & 23.32 & 21.70 & 10.65 & 7.49 & 7.83 & 9.67 & ...& 10.21 & ...& ...\\
NGC 5426/7 & 23.09 & 21.81 & 10.55 & 7.66 & 7.94 & 10.42& ...& 10.83 & ...&...\\
ZW 247.020 & 23.79 & 21.51 & 11.03 & 7.32 & 7.57 & ...& ...& 10.14 & ...&...\\
NGC 5600 & 22.70 & 21.03 & 10.09 & 6.89 & 7.13 & 9.11 & 8.97 & 10.22 & 286& 13.3\\
NGC 5653 & 23.36 & 21.63 & 10.74 & 7.49 & 7.72 & 9.23 & 9.76 & 10.49 & 240 & 9.5\\
NGC 5665 & 22.72 & 21.11 & 10.07 & 7.20 & 6.91 & 8.88 & 9.05 & 10.25 & 234 & 10.6\\
NGC 5676 & 22.97 & 21.55 & 10.42 & 7.42 & 7.68 & 9.77 & 9.96 & 10.42 & 577 & 2.9\\
NGC 5713 & 23.11 & 21.33 & 10.43 & 7.14 & 7.43 & 9.91 & 9.68 & 10.41 & 982 & 5.7\\
UGC 9618 & 24.09 & 22.51 & 11.45 & 8.35 & 8.65 & $>$10.26 &...& 10.65 & ...& ...\\
NGC 5792 & 22.78 & 21.37 & 10.15 & 7.16 & 7.50 & 10.01 & ...& 10.32 & ...&...\\
ZW 049.057 & 23.75 & 21.70 & 11.02 & 7.48 & 7.76 & ...& 9.55 & 9.71 & ...&29.5\\
NGC 5900 & 22.91 & 21.70 & 10.32 & 7.14 & 7.37 & 9.63 & ...& 9.42 & ...& ...\\
1 ZW 107 & 24.36 & 22.10 & 11.55 & 7.81 & 8.22 &...& 10.27 & 10.57 & ...& 19.2\\
NGC 5929/30 & 23.05 & 21.15 & 10.34 & 6.92 & 7.24 & 9.34 & 10.19 & 9.42 & 575 & 
8.3\\
IR 1525+36 & 24.46 & 22.20 & 11.61 & 7.80 & 8.34 &...& ...& 10.48 & ... & ...\\
NGC 5936 & 23.38 & 21.60 & 10.73 & 7.43 & 7.69 & 9.41 & 9.80 & 10.54 & 330 & 8.4\\
NGC 5937 & 23.45 & 21.49 & 10.83 & 7.33 & 7.61 & 9.21 & ...& 10.39 & ... & ...\\
NGC 5953/4 & 22.88 & 21.29 & 10.21 & 7.06 & 7.42 & 9.30 & 9.48 & 10.21 & 437 & 5.4\\
ARP 220 & 24.73 & 22.60 & 11.94 & 8.29 & 8.69 & 10.48 & 10.41 & 10.44 & 285 &33.5\\
IR 1533-05 & 23.80 & 21.91 & 11.09 & 7.71 & 8.02 & ...& ...& 10.01 & ... & ...\\
NGC 5962 & 22.78 & 21.31 & 10.19 & 7.15 & 7.43 & 9.48 & 9.46 & 10.34 &418 & 5.3\\
NGC 5990 & 23.37 & 21.42 & 10.69 & 7.25 & 7.49 & 9.23 & ...& 10.62 & ... & ...\\
NGC 6052 & 23.40 & 21.53 & 10.69 & 7.32 & 7.64 & 9.88 & ...& 10.53 & ... & ...\\
MCG+01-42-088 & 23.85 & 21.90 & 11.14 & 7.71 & 8.00 & 9.48 & ...& 10.46 & ...& ...\\
NGC 6181 & 22.96 & 21.33 & 10.37 & 7.20 & 7.44 & 9.75 & 9.79 & 10.38 & 746 & 3.9\\
NGC 7448 & 22.84 & 21.19 & 10.22 & 7.02 & 7.29 & 9.75 & 9.06 & 10.39 & 640 & 14.4\\
NGC 7469 $^p$ & 24.06 & 22.01 & 11.30 & 7.73 & 8.11 & 9.47 & 10.33 & 10.90 & 460& 9.2\\
NGC 7479 & 23.18 & 21.50 & 10.48 & 7.25 & 7.62 & 9.94 & 10.06 & 10.64 & 1141 & 2.6\\
ZW 453.062 & 23.85 & 21.78 & 11.10 & 7.56 & 7.88 & 9.46 & ...& 10.29 & ... & ...\\
NGC 7541 & 23.40 & 21.70 & 10.76 & 7.52 & 7.80 & 10.19 & 9.95 & 10.44 &726 & 6.5\\
ZW 475.056 & 24.02 & 21.90 & 11.25 & 7.66 & 7.99 & 9.74 & 10.03 & 10.44& 348 & 16.7\\
NGC 7591 & 23.53 & 21.73 & 10.84 & 7.53 & 7.83 & 10.23 & 9.96 & 10.40 & 771 & 7.6\\
NGC 7592 & 23.88 & 21.96 & 11.12 & 7.69 & 8.09 & 10.07 & 10.32 & 10.69 & 677 & 6.3\\
NGC 7674 & 23.86 & 21.10 & 11.12 & 7.84 & 8.24 & 10.33 & 10.61 & 10.83 & 891 & 3.2\\
NGC 7678 & 23.17 & 21.59 & 10.54 & 7.41 & 7.71 & 9.82 & 9.52 & 10.66 & 389 & 10.5\\
NGC 7679 & 23.53 & 21.59 & 10.79 & 7.35 & 7.70 & 9.80 & ...& 10.64 & ... & ...\\
NGC 7714 & 23.15 & 20.97 & 10.37 & 6.68 & 7.06 & 9.79 & 9.33 & 10.25 & 1739 & 11.0\\
NGC 7771 & 23.81 & 22.04 & 11.14 & 7.86 & 8.15 & 9.95 & 10.18 & 10.58 & 332 & 9.3\\
MRK 331 & 23.97 & 21.81 & 11.18 & 7.53 & 7.90 & 9.96 & 10.37 & 10.07 & 964 & 6.5\\
UGC 12914/5 & 23.32 & 21.77 & 10.70 & 7.60 & 7.89 & 10.27 & ...& 10.68 & ... & ...\\

\end{longtable}
\twocolumn
{\bf Notes to Table~\ref{massT}}\\ (1) 60$\mu$m luminosity. (2)
850$\mu$m luminosity. (3) FIR luminosity calculated by integrating the
measured SED from 40-1000$\mu$m. (4) Dust mass calculated using a
single temperature, derived from fitting the 60,100 and 850$\mu$m
fluxes. (5) Dust mass ($M_{\rm{d}}^{cold}$) calculated using a
two-component temperature fit to the data, assuming a cold
$T_{\rm{d}}=20$ K and $\beta=2$. (6) H{\sc i} refs:- Bottinelli et
al. (1990), Huchtmeier \& Richter (1989), Theureau et al. (1998) (7)
H$_2$ refs:- Young et al. (1995), Solomon et al. (1997), Sanders et
al. (1991), Chini, Kr\"{u}gel \& Lemke (1996), Maiolino et al. (1997),
Casoli et al. (1996), Lavezzi \& Dickey (1998), Sanders et al. (1986),
Sanders \& Mirabel (1985). (8) Blue luminosity calculated from
B$_{\rm{T}}$ taken from the LEDA database and corrected for galactic
extinction but not for internal extinction or inclination effects. (9)
$G_{\rm{d}}$ is the gas-to-dust ratio calculated from H{\sc i} +
H$_{2}$ and the single temperature dust mass. (10) The FIR luminosity
per unit gas mass (molecular); often used as a measure of the star
formation efficiency of a galaxy. {$^p$ NGC 7469 includes masses for IC
5283, except for H{\sc i} which is for NGC 7469 only.


%% file: table5.tex

\begin{table*}
\centering
\caption{\label{pmassT}Luminosities and masses for pairs resolved by SCUBA}
\begin{tabular}{lccccccccc}
\\[-2ex] 
\hline
\\[-2.5ex]
\multicolumn{1}{c}{Name}&\multicolumn{1}{c}{log $L_{60}$}&\multicolumn{1}{c}{log
$L_{850}$}&\multicolumn{1}{c}{log
$L_{\rm{fir}}$}&\multicolumn{1}{c}{log
$M_{\rm{d}}$}&\multicolumn{1}{c}{log $M_{\rm{d}}^{cold}$}&\multicolumn{1}{c}{log
$M_{\rm{HI}}$}&\multicolumn{1}{c}{log $M_{\rm{H_2}}$}&\multicolumn{1}{c}{log
$L_{\rm{B}}$}&\multicolumn{1}{c}{$G_{\rm{d}}$}\\
\\[-2ex]
\multicolumn{1}{c}{}&\multicolumn{1}{c}{(W Hz$^{-1}$
sr$^{-1}$)}&\multicolumn{1}{c}{(W Hz$^{-1}$
sr$^{-1}$)}&\multicolumn{1}{c}{(L$_{\odot}$}&\multicolumn{1}{c}{(M$_{\odot}$)}&\multicolumn{1}{c}{(M$_{\odot}$)}&\multicolumn{1}{c}{(M$_{\odot}$)}&\multicolumn{1}{c}{(M$_{\odot}$}&\multicolumn{1}{c}{(L$_{\odot}$)}&\multicolumn{1}{c}{}\\
\\[-2.5ex]
\hline
\\[-2.5ex]
IC 563 & 23.28 & 21.78 & 10.65 & 7.59 & 7.92 &\shortstack{10.10\\[-2ex] } & ... & 10.09 & ...\\
IC 564 & 23.21 & 21.86 & 10.66 & 7.73 & 8.00 &  & ... & 10.35& ...\\
\\[-2ex]
NGC 3994 & 23.04 & 21.22 & 10.29 & 7.07 & 7.33 & 9.75 & ... & 9.86 & ...\\
NGC 3995 & 23.11 & 21.34 & 10.16 & 7.14 & 7.47 & 10.09 & ... & 10.33 & ...\\
\\[-2ex]
NGC 5257 &... & 21.92 & ... & 7.69 & ... &10.27 & ... & 10.69 & ...\\
NGC 5258 & ... &22.09 & ... & 7.85 & ... &10.23 & ... & 10.71 & ...\\
\\[-2ex]
NGC 5394 & ... &21.14 & ... & 7.02 & ... & 9.95 & ... & 10.06 & ...\\
NGC 5395 & ... &21.59 & ... & 7.47 & ... & 10.13 & ... & 10.49 & ...\\
\\[-2ex]
NGC 5426 & 22.56 & 21.51 & 10.06 & 7.36 & 7.64 & 10.16 & ... & 10.22 & ...\\
NGC 5427 & 22.95 & 21.51 & 10.38 & 7.36 & 7.63 & 10.08 & ... & 10.62 & ...\\
\\[-2ex] 
UGC 9618 (N) & ...& 22.32 & ... & 8.17 & ... &\shortstack{$>$10.26\\[-2ex] } & ... & 10.59&...\\
UGC 9618 (S)  & ...& 22.07 & ... & 7.91 & ... &   & ... & 10.28 & ...\\
\\[-2ex]
NGC 5929 & ... &20.41 & ... & 6.19 & ... & 8.93 & 9.06 & 9.61 & 1291\\
NGC 5930 & ... &21.05 & ... & 6.83 & ... & 9.08 & 9.14 & 9.85 & 373\\
\\[-2ex]
NGC 5953 & ... &21.07 & ... & 6.84 & ... & 9.06 & 9.27 & 9.70 & 435\\
NGC 5954 & ... &20.90 & ... & 6.67 & ... & 8.93 & 9.05 & 9.54 & 423\\
\\[-2ex]
NGC 7469 & ... &21.87 & ... & 7.52 & ... & 9.48 & 10.11 & 10.62 & 480\\
IC 5283 & ... &21.44 & ... & 7.27 & ... & ... & 9.94 & 9.78 & ...\\ 
\\[-2ex]
UGC 12914 & ...&21.61 & ... & 7.44 & ... & 9.99 & ... & 10.49 & ...\\
UGC 12915 & ...&21.69 & ... & 7.52 & ... & 9.95 & ...& 10.13 & ...\\
\\[-2.5ex]
\hline
\\[-2.5ex]
\multicolumn{7}{p{34em}}{\small{Columns have the same meanings as in Table~\ref{massT}}}\\
\end{tabular}
\end{table*}

%% file: table6.tex

\begin{table}
\centering
\caption{\label{convT} Conversion factors used for CO data}
\begin{tabular}{lclcc} 
\\[-2ex]
\hline
\\[-2.5ex]
\multicolumn{1}{c}{Telescope}&\multicolumn{1}{c}{Beam
size}&\multicolumn{1}{c}{Scale}&\multicolumn{1}{c}{Conversion}&\multicolumn{1}{c}{Refs}\\
\multicolumn{1}{c}{}&\multicolumn{1}{c}{\small{$^{\prime\prime}$}}&\multicolumn{1}{c}{}&\multicolumn{1}{c}{(Jy
K$^{-1}$)}&\multicolumn{1}{c}{}\\
\\[-2.5ex]
\hline
\\[-2.5ex]
FCRAO & 45 & $T^{\ast}_A$ & 42 & a\\
      &    & $T^{\ast}_R$ & 31.5 & d, h, i\\
SEST & 45 & $T_{mb}$ & 19 & b\\
IRAM & 22 & $T_{mb}$ & 4.5 & c\\
NRAO & 55 & $T^{\ast}_R$ & 35 & d, e, i\\
     &    & $T_{mb}$ & 30 & f, g\\
\\[-2.5ex]
\hline
\\[-2.5ex]
\multicolumn{5}{p{26em}}{\small {Reference Key :- (a) Young et al. (1995), (b) Chini et
al. (1996), (c) Solomon et al. (1997), (d) Sanders et
al. (1991), (e) Maiolino et al. (1997), (f) Casoli et
al. (1996), (g) Lavezzi \& Dickey (1998), (h) Sanders
et al. (1986), (i) Sanders \& Mirabel (1985)}}\\
\end{tabular}
\end{table}

%% file: table7.tex

\begin{table}
\centering
\caption{\label{lfT}Local luminosity function at 850$\mu$m}
\begin{tabular}{cccc}
\\[-2ex] 
\hline
\\[-2.5ex]
\multicolumn{1}{c}{log $L_{850}$}&\multicolumn{1}{c}{$\phi(L)$}&\multicolumn{1}{c}{$\sigma_{\phi}$}\\
\multicolumn{1}{c}{(W Hz$^{-1}$
sr$^{-1}$)}&\multicolumn{1}{c}{(Mpc$^{-3}$
dex$^{-1}$)}&\multicolumn{1}{c}{(Mpc$^{-3}$
dex$^{-1}$)}\\
\\[-2.5ex]
\hline
\\[-2.5ex]
21.04 & 2.43e-3 & 7.69e-4 &  \\
21.28 & 1.33e-3 & 3.32e-4 & \\
21.52 & 7.00e-4 & 1.37e-4 &\\
21.76 & 2.13e-4 & 4.65e-5 &\\
22.00 & 6.02e-5 & 1.38e-5 &\\
22.24 & 2.34e-5 & 7.8e-6 &\\
22.48 & 1.60e-6 & 9.2e-7 &\\
\\[-2ex]
\hline
\\[-2ex]
\multicolumn{1}{c}{$\alpha$}&\multicolumn{1}{c}{$L_{\ast}$}&\multicolumn{1}{c}{$\phi_{\ast}$}&\multicolumn{1}{c}{$\chi^{2}_{\nu}$}\\
\multicolumn{1}{c}{}&\multicolumn{1}{c}{(W Hz$^{-1}$
sr$^{-1}$)}&\multicolumn{1}{c}{(Mpc$^{-3}$
dex$^{-1}$)}&\multicolumn{1}{c}{}\\
\\[-2ex]
\hline
\\[-2ex]
$-2.18^{+0.22}_{-0.24}$ &
$8.3^{+2.5}_{-2.0}\times 10^{21}$ & $2.9^{+4.1}_{-1.5}\times 10^{-4}$ & 
0.68\\
\\[-2ex]
\hline
 & & \\
\multicolumn{3}{l}{Single temperature dust mass function}\\
\\[-2.5ex]
\hline
\\[-2.5ex]
\multicolumn{1}{c}{log $M_{\rm{d}}$}&\multicolumn{1}{c}{$\phi(M)$}&\multicolumn{1}{c}{$\sigma_{\phi}$}\\
\multicolumn{1}{c}{(M$_{\odot}$)}&\multicolumn{1}{c}{(Mpc$^{-3}$
dex$^{-1}$)}&\multicolumn{1}{c}{(Mpc$^{-3}$ dex$^{-1}$)}\\
\\[-2.5ex]
\hline
\\[-2.5ex]
6.78 & 2.00e-3 & 6.68e-4 & \\
7.02 & 1.05e-3 & 3.16e-4 &\\
7.26 & 1.01e-3 & 2.10e-4 &\\
7.50 & 4.82e-4 & 9.28e-5 &\\
7.75 & 1.32e-4 & 2.75e-5 &\\
7.99 & 2.27e-5 & 8.0e-6 &\\
8.23 & 1.58e-6 & 9.1e-7 &\\
\\[-2ex]
\hline
\\[-2ex]
\multicolumn{1}{c}{$\alpha$}&\multicolumn{1}{c}{$M_{\ast}$}&\multicolumn{1}{c}{$\phi_{\ast}$}&\multicolumn{1}{c}{$\chi^{2}_{\nu}$}\\
\multicolumn{1}{c}{}&\multicolumn{1}{c}{(M$_{\odot}$)}&\multicolumn{1}{c}{(Mpc$^{-3}$
dex$^{-1}$)}&\multicolumn{1}{c}{}\\
\\[-2ex]
\hline
\\[-2ex]
$-1.23^{+0.44}_{-0.59}$ &
$2.50^{+0.95}_{-0.82}\times 10^7$ & $1.64^{+0.69}_{-0.48}\times
10^{-3}$ & 0.56\\
\\[-2ex]
\hline
 & & \\
\multicolumn{3}{l}{Cold component dust mass function}\\
\\[-2.5ex]
\hline
\\[-2.5ex]
\multicolumn{1}{c}{log $M_{\rm{d}}^{cold}$}&\multicolumn{1}{c}{$\phi(M)$}&\multicolumn{1}{c}{$\sigma_{\phi}$}\\
\\[-2.5ex]
\hline
\\[-2.5ex]
7.10 & 2.30e-3 & 7.67e-4 & \\
7.34 & 1.18e-3 & 3.04e-4 &\\
7.59 & 7.91e-4 & 1.61e-4 &\\
7.83 & 2.74e-4 & 5.99e-5 &\\
8.08 & 7.63e-5 & 1.63e-5 &\\
8.32 & 2.57e-5 & 8.13e-6 &\\
8.57 & 1.56e-6 & 9.02e-7 &\\
\\[-2ex]
\hline
\\[-2ex]
\multicolumn{1}{c}{$\alpha$}&\multicolumn{1}{c}{$M^{cold}_{\ast}$}&\multicolumn{1}{c}{$\phi_{\ast}$}&\multicolumn{1}{c}{$\chi^{2}_{\nu}$}\\
\\[-2ex]
\hline
\\[-2ex]
$-1.91^{+0.21}_{-0.24}$ & $
8.5^{+2.2}_{-1.9}\times 10^7$ & $ 4.9^{+3.3}_{-2.2}\times 10^{-4}$ & 0.62\\
\\[-2.5ex]
\hline
\end{tabular}
\end{table}


%% file: gas.tex


\section{DISCUSSION}

\subsection{Correlations}
\subsubsection{Dust temperature}

\begin{figure}
 \vspace{6cm}
 \includegraphics{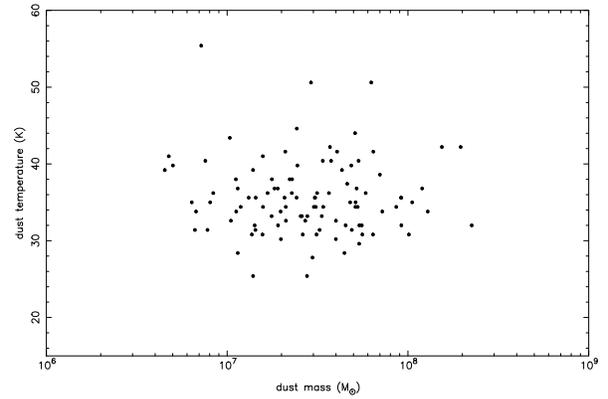}
\caption{\label{md2tdF} Plotting dust temperature versus dust mass shows no
correlation. This means that $L_{60}$--$T_{\rm{d}}$ relationship
(Fig.~\ref{60TF}) is not a function of dust mass.}
\end{figure}

\begin{figure}
 \vspace{6cm}
 \includegraphics{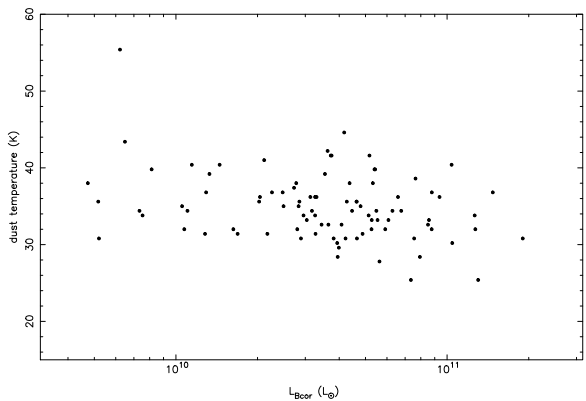}
\caption{\label{lb2tdF} A plot of dust temperature against blue luminosity
shows an anti-correlation ($r=0.2$). This is quite surprising as
initially it was thought that the $L_{60}$--$T_{\rm{d}}$ correlation
was due to galaxy mass (i.e. larger galaxies have more 60$\mu$m
luminosity and therefore higher dust temperatures).}
\end{figure}

\begin{figure}
 \vspace{7.7cm}
 \includegraphics{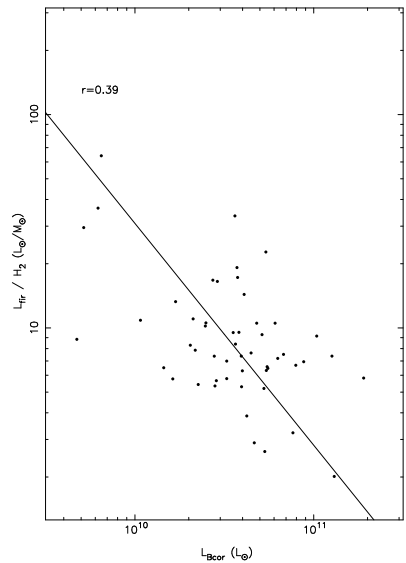}
\caption{\label{sfe2lbF} A stronger anti-correlation is seen when the dust
temperature from Fig.~\ref{lb2tdF} is replaced by a measure of the star
formation efficiency ($L_{\rm{fir}}/M_{\rm{H_2}}$). This is the
explanation for the weak dependence on dust temperature in Fig.~\ref{lb2tdF},
as $L_{\rm{fir}}$ is a function of $T_{\rm{d}}$. This figure suggests
that smaller galaxies have larger star formation efficiencies, a trend
first noticed by Young (1999).}
\end{figure}

Luminosity is often strongly correlated with mass in astrophysical
situations, and so at first sight the $L_{60}-T_{\rm{d}}$ correlation
(Fig.~\ref{60TF}) suggests that the more luminous systems are
hotter. We can test this by plotting dust temperature versus dust mass
and blue luminosity, often used as a measure of the mass of a
galaxy. In the first case (Fig.~\ref{md2tdF}) there is no significant
correlation (see Table~\ref{fitsT} for coefficients), so there is no
tendency for galaxies with lots of dust to be hotter, and in the
second case (Fig.~\ref{lb2tdF}) there is an inverse correlation. The
$L_{60}-T_{\rm{d}}$ correlation is therefore most naturally explained
by the sensitivity of the 60$\mu$m flux to the dust temperature. If,
for whatever reason (and not because of its mass) the dust is hotter,
this will greatly increase the 60$\mu$m luminosity. What is rather
more interesting (and perhaps surprising) is the anti-correlation of
the corrected $L_{\rm{B}}$ versus $T_{\rm{d}}$
(Fig.~\ref{lb2tdF}). The significance is at the 98 per cent level,
making it the least secure of the correlations we present here and the
probable link is with the star formation efficiency rather than
dust temperature itself. Using $L_{\rm{fir}}/M_{\rm{H_2}}$ as a
measure of star formation efficiency, Young (1999) has recently found
that the star formation efficiency of galaxies in many different types
of environment decreases as galaxy size increases (as measured by the
optical linear diameter $D_{25}$). If we now plot
$L_{\rm{fir}}/M_{\rm{H_2}}$ against $L_{\rm{B}}$ for our galaxies a
much better correlation is seen than with $T_{\rm{d}}$, indicating
that this is where the true dependence lies (see Fig.~\ref{sfe2lbF}
and Table~\ref{fitsT}). The explanation Young gives for this
relationship is that larger galaxies have flatter rotation curves over
more of their disks and so experience more shear. She suggests that
the effect of this would be to reduce the ability of molecular clouds
to form stars by increasing the turbulence within them, and to dampen
the effects of star-formation triggers.

\input{table8}

\subsubsection{\label{7714}Gas and dust masses}

\begin{figure*}
 \vspace{8.5cm}
 \includegraphics{Fig14a.ps}
 \includegraphics{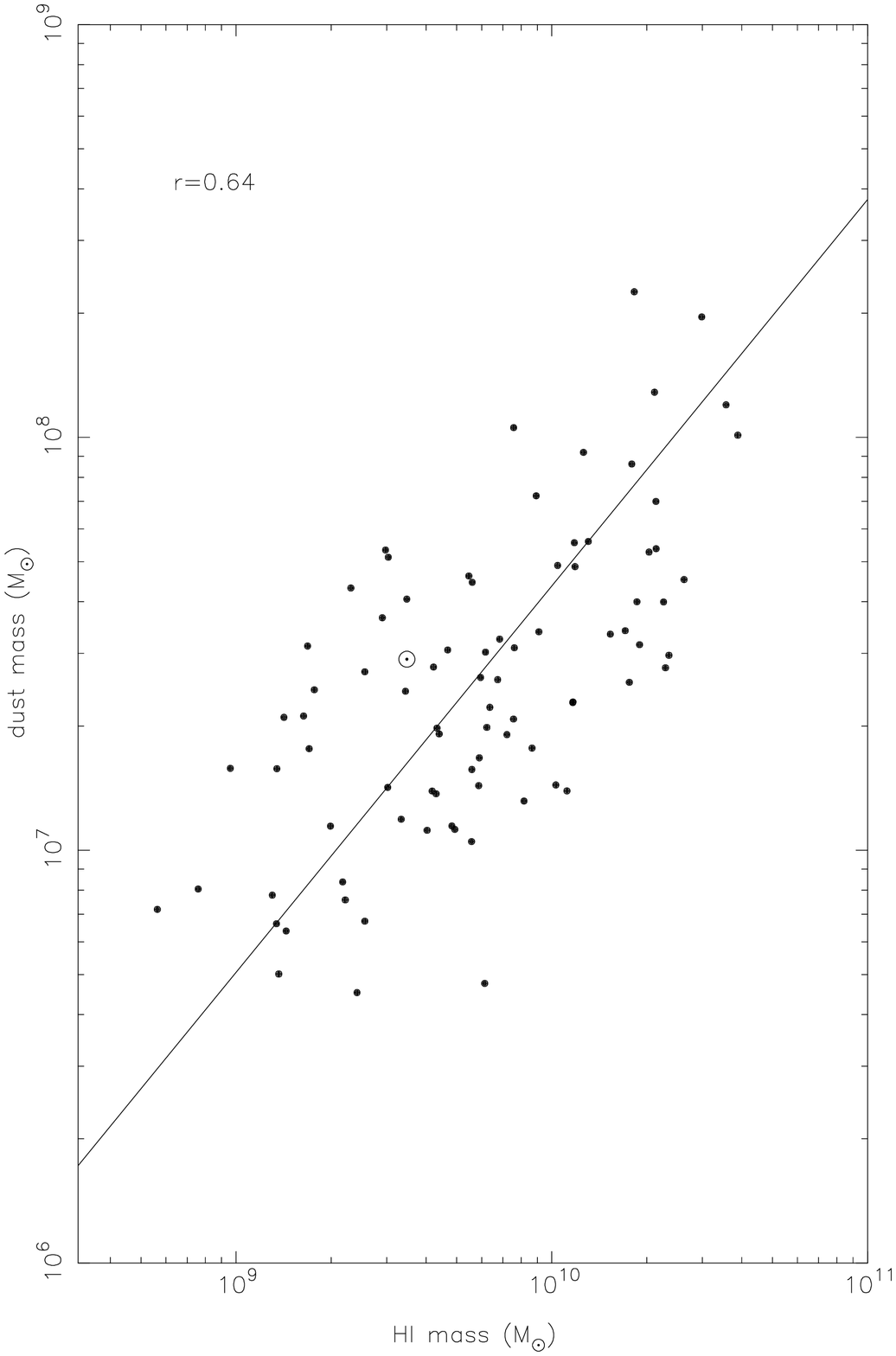}
 \includegraphics{Fig14c.ps}
\caption[a]{\label{d2gF} a) Dust mass versus H$_2$ mass, a solar symbol
indicates the Milky Way values (Sodroski et al. 1994). The line is the
best least squares fit (see Table~\ref{fitsT} for fit parameters).\\
b) Dust mass versus H{\sc i} mass, the line indicates the best least
squares fit to the data. There is much more dispersion than for
Fig.~\ref{d2gF}a.\\ c) Dust mass versus H$_2$ + H{\sc i} mass.}
\end{figure*}

We compared the dust masses, derived from the submillimetre fluxes
using a single temperature, to the H$_2$, H{\sc i} and H$_2$ + H{\sc
i} masses as shown in Figure~\ref{d2gF}a-c. The strongest correlation
is with the molecular gas which suggests that the dust is primarily
found in molecular clouds. The larger scatter on the H{\sc i} plots
may be because we are using global H{\sc i} values which may include
gas at very large radii where there would be much less dust than in
the inner disk. Devereux \& Young (1990) found that if
they only used the H{\sc i} in the inner disk, in addition to the
molecular gas, then the correlation between gas and dust was
improved. We cannot, however, investigate the spatial connection between
dust and atomic gas in the region they both occupy as we do not yet
have H{\sc i} maps. We repeated the process using the `cold dust masses' but
found no differences in the correlations or slopes of the plots and so
have not included these.

We would like to try to address the reasons behind the tightness of
the relationship between dust and molecular gas. The scatter in this
plot (Fig.~\ref{d2gF}a) is very small, the r.m.s dispersion about the
line of best fit is only $\sim 60$ per cent. Some of this is due to
observational errors; if we assume 30 per cent errors in the fluxes and
subtract this in quadrature from the overall value then we are left
with 52 per cent as the intrinsic scatter. There are some
observational-type errors which are hard to account for, such as
aperture corrections and the probability that our assumption of a
single dust temperature is wrong, but the small scatter does suggest
that any aperture effects must be similar for all the galaxies and
that if our assumption about temperature is wrong then it must be
wrong in the same way for all of the galaxies. This is borne out be
the similarity of the relationship when the `cold dust masses' are
used.

In order to determine the physical meaning of the 50 per cent scatter
we will review the current views on the formation of dust and CO and
the expected dependences on metallicity.

There are no firm conclusions about how dust grains form. Suggested
sites of formation range from post-AGB stars to supernova remnants
(Whittet 1992). The grains are thought to grow in the darkest molecular
clouds by accretion of icy mantles and are probably destroyed by
sputtering in the diffuse ISM, after the dispersion of the molecular
cloud by supernova shocks and HII regions. Despite the complications
in determining how and where the various stages of dust evolution
occur, there is some evidence that the fraction of metals bound up in
dust grains is a constant, both from observations of nearby galaxies
(Issa et al. 1990) and from observations in our own Galaxy, where the
carbon and oxygen abundances (the main constituents of dust) are
constant over a wide range ISM densities (Cardelli et al. 1996; Meyer
et al. 1998). This suggests that $M_{\rm{d}}
\propto Z M_{g}$ where $M_g$ is the total mass of gas and $Z$ is the metallicity.

The formation of CO is better understood than that of dust and is
believed to occur through a network of gas-phase reactions in the
giant molecular clouds (GMCs). CO will form readily, the rate
increasing with density, but is easily destroyed by UV radiation. In
order to remain intact, the CO needs to be shielded by dust grains
and/or large column densities of H$_2$. Assuming that the dominant
mass in the GMCs is in the form of H$_2$, then since CO forms from the
C and O available in the gas-phase, (and O is always more abundant
than C in the ISM), a metallicity dependence is suggested where
$L_{\rm{CO}} \propto Z M_{\rm{H_2}}$. However, since most of the
shielding required to form the CO is produced by dust rather than
H$_2$, the true metallicity dependence may be even higher than
this. When there is insufficient dust in a molecular cloud to provide
adequate shielding from the UV radiation field, the volume of the
cloud occupied by CO may not be as large as that occupied by H$_2$,
because H$_2$ is more efficient at self-shielding. This is the likely
case in low metallicity systems such as the LMC or where the volume
and/or column densities are low. The possibilities that CO and H$_2$
may not always be co-extensive, and that there may be a dependence of
CO luminosity on metallicity, can lead to problems with the
application of the $X$ factor to CO observations in order to determine
the mass of H$_2$. For systems with significantly lower densities
and/or metallicities a larger conversion may be needed relative to
galaxies such as the Milky Way. Observations do indicate a dependence
of $X$ on metallicity for low metallicity systems (Wilson 1995) with
$X/X_G= (Z/Z_{\odot})^{-0.7}$ for systems with less than solar
metallicity and staying roughly constant for higher than solar
metallicities (Frayer \& Brown 1997). This observed dependence is less
steep than one would have imagined from the above simplistic
arguments. It could be argued that for metallicities above some
critical value, where there is sufficient shielding, the $X$ factor
may show less dependence on metallicity, as the abundance of CO
relative to H$_2$ depends on the availability of C and O and not on
dust any longer. While the formation of CO should still be greater for
higher metallicities, H$_2$ formation is also dependent on dust grains
as catalysts so it is possible that this could cause the $X$ factor to
saturate : more metals lead to more CO, but they also lead to more
dust and thus more H$_2$ formed from atomic hydrogen, producing the
observation that the $X$ factor is constant with metallicity above a
threshold value. Theory does predict significant depletion of CO onto
dust grains (forming mantles) for favourable conditions in the cloud
which would make $X$ smaller with increasing metallicity, possibly
offsetting any trend for an increase but in the Milky Way this
depletion is only observed to be 5 to 40 per cent (Whittet 1992), much
less than predicted, indicating that there is still much which is not
understood about the interaction of dust and molecules in the GMCs.

We do not have metallicity measurements for our galaxies, but the
range of metallicities of spiral galaxies of similar absolute
magnitudes is 2--3 (Henry \& Worthey 1999). Thus the small scatter in
the H$_2$--dust mass diagram strongly suggests that the metallicity
dependence of dust mass and of the CO $X$ factor is very similar. An
additional test of this will be extending our study to galaxies likely
to have a larger range of metallicity.

\subsubsection{Star formation efficiency}

\begin{figure}
 \vspace{8cm}
 \includegraphics{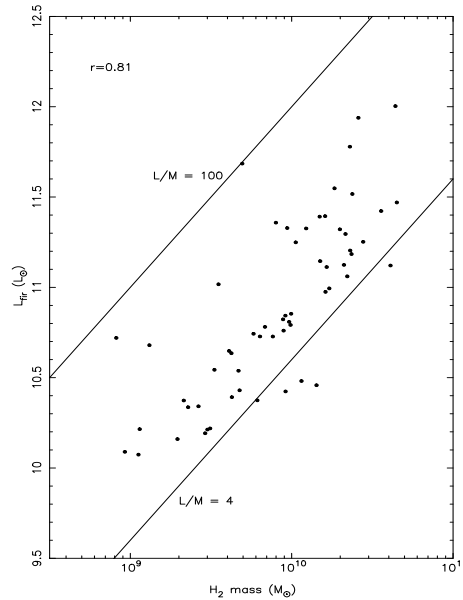}
\caption{\label{lfir2h2F} FIR luminosity versus H$_2$ mass. The
quantity $L_{\rm{fir}}/M_{\rm{H_2}}$ is taken as a measure of the star
formation efficiency of the galaxy. The galaxies lying above the trend
are mostly interacting or with suspected active nuclei. Our Galaxy is
believed to lie on the L/M$\sim4$ line and so most of the galaxies in
our sample would appear to be more efficient than the Milky Way.}
\end{figure}

The ratio of $L_{\rm{fir}}/M_{\rm{H_2}}$ is often used as a measure of
the star formation efficiency of a galaxy since $L_{\rm{fir}}$ is
believed to trace the star formation rate and so this ratio gives the
star formation rate per unit gas mass. Figure~\ref{lfir2h2F} shows the
strong correlation between molecular gas and $L_{\rm{fir}}$ and gives
some idea of the range of $L_{\rm{fir}}/M_{\rm{H_2}}$ values for these
galaxies. The L/M = 4 line represents `normal' galaxies such as the
Milky Way (Scoville \& Good 1989) and most of the galaxies in our
sample show elevated star formation efficiencies relative to the Milky
Way. Many of the galaxies lying above the general trend (SFE$\geq 20$)
are often interacting or have suspected active nuclei (e.g. IR 0857+39
with SFE $\sim 100$). Values of $L_{\rm{fir}}/M_{\rm{H_2}}$ are
given in Table~\ref{massT}.

\subsection{\label{g2dS}Gas to dust ratios}

The average gas-to-dust ratios (using the single temperature dust
masses) for the three components are
$M_{\rm{H_2}} /M_{\rm{d}} = 293\pm19$, $M_{\rm{HI}} /M_{\rm{d}}
=304\pm24$ and $M_{\rm{H_2 + HI}} /M_{\rm{d}} = 581\pm43$, where the
$\pm$ indicates the error in the mean. In future, we will refer to
$M_{\rm{H_2 + HI}} /M_{\rm{d}}$ as the gas-to-dust ratio $G_{\rm{d}}$
and the values are listed in Table~\ref{massT}. In order to compare our
$G_{\rm{d}}$ with the Galactic value, we must be careful that we
account for the differences in $\kappa_{d}(\nu)$ and $X$ used by
others and ourselves. The gas-to-dust ratio for the Milky Way was
measured by Sodroski et al. (1994), using {\em COBE\/} data at 140 and
240$\mu$m, therefore in order to compare our measurement with theirs
we must also scale the $\kappa_{d}(240)$ value of the opacity
coefficient they used to 850$\mu$m, using an assumed value for
$\beta$. Sodroski et al. use a $\kappa_{d}(240)=0.72\,
\rm{m}^{2}\,\rm{kg}^{-1}$; taking this to 850$\mu$m gives
$\kappa_{d}(850)=0.2-0.057\,\rm{m}^{2}\,\rm{kg}^{-1}$ for
$\beta=1-2$. Comparing this to our value of 0.077 m$^2$ kg$^{-1}$ and
accounting for the differences in $X$ factors (theirs being 0.71 times
ours) gives a range of Galactic $G_{\rm{d}}$ values of 85--302 for a
$\beta$ of $1\rightarrow 2$. The middle of the range at $\beta=1.5$ is
$G_{\rm{d}}=160$ and this is the nominal value we will make
comparisons with. It can now be seen that our average gas-to-dust
ratio of 581 is more than a factor of three larger than the Galactic
value of 160 (at the least a factor of 2 larger if $\beta=2$). Of
course, the measurement made by Sodroski et al. is a very difficult
one to make, but it is supported by arguments based on element
abundances and depletions which predict a ratio of $\sim 130$ (Eales
et al. in prep). The Galaxy has a FIR luminosity of
$1.1\times 10^{10}\, \rm{L_{\odot}}$ and a dust mass of $\sim
2.9\times 10^{7}\,\rm{M_{\odot}}$ (Sodroski et al. 1994) which assumes
the same value for $\kappa_{d}(\nu)$ that gives a $G_{d}=160$. This
places the Galaxy at the bottom end of the range of luminosities and
dust masses found in this IR bright sample. Previous determinations
of the gas-to-dust ratios in IR luminous objects using {\em IRAS\/}
data have always been high $\sim 1000$ (Young et al. 1989; Devereux \&
Young 1990; Sanders et al. 1991) which has been attributed to the lack
of sensitivity of {\em IRAS} to cold dust. It would seem that our
submillimetre observations have reduced this tendency, giving a lower
gas-to-dust value, but when the gas-to-dust ratios are re-calculated
using the `cold dust masses' we find a further reduction to
$M_{\rm{H_2}} /M_{\rm{d}}^{cold} = 143\pm9$, $M_{\rm{HI}}
/M_{\rm{d}}^{cold} =158\pm14$ and $M_{\rm{H_2 + HI}}
/M_{\rm{d}}^{cold} = 293\pm24$. This is now just within the range of
values we calculated for the Galaxy, and in good agreement if the
$\beta=2$ case is taken. We perform the calculations with the `cold
masses' simply to illustrate the effect of overlooking this
component. A slightly colder temperature (say 15--18 K) would increase
the dust masses by a larger amount, further reducing the observed
gas-to-dust ratios and so giving better agreement with the Galactic
values.

The gas-to-dust ratios calculated using the single temperature dust
masses vary from 1700 to 200, the large range being mostly due to the
variations in $M_{\rm{HI}} /M_{\rm{d}}$ as discussed in
Section~\ref{7714}. Some of the larger values of $M_{\rm{H_2}}
/M_{\rm{d}}$ could be due to the large angular sizes of some of the
nearer galaxies, meaning that our dust masses are underestimated when
compared to the CO fluxes produced from multi-positional single dish
measurements (this particularly applies to NGC 772 and NGC 7479);
however this is not always the case as there are some galaxies of
small angular size which have high $M_{\rm{H_2}} /M_{\rm{d}}$ ratios
but more normal $M_{\rm{HI}} /M_{\rm{d}}$, (e.g. IR 0335+15, UGC 4881
and Mkn 331). These galaxies are interacting or have Seyfert nuclei
and typically have high dust temperatures, which may suggest that the
lack of apparent dust mass is simply due to using a high temperature
when calculating $M_{\rm{d}}$ and not accounting for any colder
dust. High temperatures cannot be the sole explanation, however, as
other galaxies with high $T_{\rm{d}}$ and signs of
interaction/activity have rather low values of $M_{\rm{H_2}}
/M_{\rm{d}}$ (e.g. Arp 220). It could be that an enhanced
$M_{\rm{H_2}} /M_{\rm{d}}$ is a stage that some merging/starburst
galaxies pass through, either by an enhancement of CO relative to
dust, or a decrease in dust emission. A change in the $X$ factor,
causing the H$_2$ to be overestimated could also be responsible (this
has been suggested for many extreme active/starburst nuclei (Solomon
et al. 1997)). It is also possible that CO which was depleted from the
gas phase onto grain mantles in dark clouds could, at a certain stage
in the starburst evolution, become vaporized back into the gas phase,
giving a higher CO flux while keeping the atomic, dust and H$_2$
content the same. One particular case which deserves a mention here is
NGC 7714 which has $M_{\rm{HI}}/M_{\rm{d}} \sim 1300$, causing its
abnormally high gas-to-dust ($G_{\rm{d}}$) value of 1700 (see column 9
of Table~\ref{massT}). NGC 7714 is interacting with NGC 7715 although
they are widely spaced enough (1.9 arcmin) that only NGC 7714 was
observed with SCUBA and the {\em IRAS\/} flux is due entirely to NGC
7714 (Surace et al. 1993). The H{\sc i} maps of Smith et al. (1997)
show rings and bridges of gas connecting the two galaxies, so it is
probable that some of the H{\sc i} listed for NGC 7714 in the single
dish measurement (Table~\ref{massT}) is really associated with the
whole region and cannot be compared with what we observed. The amount
of H{\sc i} contained in the 50 $\times$ 50 arcsec region centred on
NGC 7714 (similar to the area of the dust emission) is given as only
$1.7\times10^9$ M$_{\odot}$ compared to $7\times10^9$ M$_{\odot}$ for
the whole system. Using this lower value would give a more reasonable
value of 355 for $M_{\rm{HI}} /M_{\rm{d}}$, bringing $G_{\rm{d}}$ down
to 802.

One further caveat to the whole $M_{\rm{H_2}}/M_{\rm{d}}$ business is
the large discrepancies in the CO fluxes between single-dish
measurements and interferometer mappings. Maps have been found for
four of our galaxies (Sanders et al. 1988; Yun \& Hibbard 1999;
Laine et al. 1999) and always the flux is considerably less ($\leq
1/2$) than the single-dish measurement (Young et al. 1995). This is
usually explained by resolution effects as these interferometer
measurements are insensitive to structures larger than $\sim 30$
arcsec but even in sources unresolved by the interferometer at $cz\geq
10,000$ km s$^{-1}$ (which are not expected to have large scale angular
structure), the difference in flux is still as large. If we used the
interferometer measurements for NGC 520, NGC 7469, NGC 7479 and UGC
4881, the corresponding H$_2$ masses would decrease by factors of 2,
2, 4, and 2 and bringing the $M_{\rm{H_2}}/M_{\rm{d}}$ ratios down to
229, 195, 163 and 260 respectively.


%% file: table8.tex

\begin{table*}
\caption{\label{fitsT} Parameters for fits and correlations}
\begin{tabular}{llcrrrr} 
\\[-2ex]
\hline
\\[-2.5ex]
\multicolumn{1}{c}{$y$}&\multicolumn{1}{c}{$x$}&\multicolumn{1}{c}{Number}&\multicolumn{1}{c}{$r_{s}$}&\multicolumn{1}{c}{significance}&\multicolumn{2}{c}{linear
fit: $y=mx+c$}\\
\multicolumn{5}{l}{}&\multicolumn{1}{c}{$m$}&\multicolumn{1}{c}{$c$}\\
\\[-2.5ex]
\hline
\\[-2.5ex]
$T_{\rm{d}}$ & log $L_{60}$ & 104 & 0.44 & $2.5e-6$ & $4.57\pm 0.82$ &
$-71.9\pm 19.4$\\
$T_{\rm{d}}$ & log $M_{\rm{d}}$ & 104 & $-0.02$ & 0.8 & &\\
$T_{\rm{d}}$ & log $L_{\rm{B}}$ $^c$ & 91 & $-0.25$ & 0.02 & & \\
$\beta$ & $T_{\rm{d}}$ & 104 & 0.44 & $3.7e-6$ & $-0.0184\pm 0.004$ &
$1.95\pm 0.13$\\
log $M_{\rm{d}}$ &log $M_{\rm{H_2}}$ & 59 & 0.89 & $6.2e-21$ & $0.82\pm
0.04$ & $-0.67\pm 0.38$\\
log $M_{\rm{d}}$ &log $M_{\rm{HI}}$ & 84 & 0.64 & $3.6e-11$ & $0.94\pm 0.09$
& $-1.81\pm 0.68$\\
log $M_{\rm{d}}$ &log $M_{\rm{H_{2}+HI}}$ & 47 & 0.86 & $7e-15$ &
$0.96\pm 0.06$ & $-2.32\pm 0.53$\\
log $L_{\rm{fir}}$ &log$M_{\rm{H_2}}$ & 59 & 0.81 & $1.4e-14$ &
$1.14\pm 0.09$ & $-0.39\pm 0.78$\\
log $L_{\rm{fir}}/M_{\rm{H_2}}$ & log $L_{\rm{B}}$ $^c$ & 52 & $-0.39$ &
$3.9e-4$ & $-1.04\pm 0.20 $ & $11.89\pm 0.59$\\
\\[-2.5ex]
\hline
\\[-2.5ex]
\multicolumn{7}{p{35em}}{\small {\sc notes} -- {\it Column(4)}--
Spearman rank correlation coefficient. {\it Column(5)}-- probability
that $x$ and $y$ are unrelated. $c$ :- luminosity corrected for galactic and
internal extinction and for inclination effects.}\\

\end{tabular}
\end{table*}

%% file: conc.tex

\parindent = 1em

\section{CONCLUSIONS}

We have undertaken the first statistical survey of the local universe
with the SCUBA bolometer array on the JCMT. We present here the
initial results from the first of the samples we are surveying: a
sample of 104 galaxies selected at 60$\mu$m from the {\em IRAS\/}
Bright Galaxy Sample.\\

(i) The 60, 100 and 850$\mu$m fluxes are well fitted by single
temperature SEDs. The mean and standard deviation (S.D.) in the
best-fitting dust temperature is $\overline{T_{\rm{d}}} = 35.6\, \pm\, 4.9$ K with the mean and S.D. for the
emissivity index being $\overline{\beta} = 1.3\, \pm\, 0.2$.  We
do not however, rule out the possibility of colder dust and a steeper
emissivity. If $\beta = 2$ and we assume there is a cold dust
component with a temperature of 20 K, we then obtain dust masses a
factor of 1.5--3 higher. The 450$\mu$m data obtained for 30 per cent
of the galaxies may eventually constrain the submillimetre emissivity index
and therefore the presence of a cold component.

(ii) We have presented the first direct measurements of the
submillimetre luminosity and dust mass functions. They are well fitted
by Schechter functions, in contrast to the {\em IRAS\/} 60$\mu$m LFs
(Lawrence et al. 1986; Rieke \& Lebofsky 1986) which do not have the
exponential cut-off required by the Schechter function. The slope of
the 850$\mu$m LF at low luminosities is steeper than $-2$ implying that the
LF must flatten at lower luminosities than were probed by this
survey. The optically selected sample currently being observed will
take the submillimetre LF to lower luminosities and constrain the
`knee' of the LF. The shape of the dust mass function is affected by
the model assumed for the temperature distribution.

(iii) We have shown that a simple extrapolation of the {\em IRAS\/}
60$\mu$m LF to 850$\mu$m using a single dust temperature and
emissivity index does not reproduce the measured submillimetre LF, both in
terms of normalisation and shape. 

(iv) A correlation was found between the fitted dust temperature and 
60$\mu$m luminosity. This is best explained by the sensitivity of
the 60$\mu$m flux to temperature (due to its position on the Wien side
of the grey-body curve) rather than by a dependence on galaxy
mass. Accounting for this temperature dependence when extrapolating
the 60$\mu$m LF to submillimetre wavelengths produces a much better
match to the observed 850$\mu$m LF.

(v) If there is a population of cold ($T_{\rm{d}} < 25$ K) galaxies
which are also luminous submillimetre sources, then our submillimetre
LF is likely to be biased, as these objects would not have appeared in
the original {\em IRAS\/} bright galaxy sample which was selected at
60$\mu$m. The question of a missing population will be addressed by
the optically selected sample. 

(vi) We find an average value for the gas-to-dust ratio ($G_{\rm{d}}$)
of $581 \pm 43$ where the gas mass is taken to be $M_{\rm{HI}} +
M_{\rm{H_2}}$. This is lower than previous values determined using
{\em IRAS\/} fluxes alone ($\sim 1000$) indicating that the
submillimetre is a better place to measure the dust mass. It is still,
however, a factor of $\sim 3.5$ higher than the value obtained for the
Galaxy (Sodroski et al. 1994). Using `cold dust masses', calculated on
the basis of there being a $20 \rm{K}$ component in addition to the warm
dust, $G_{\rm{d}}$ was reduced to 293. This supports the idea that the
previous large discrepancies in $G_{\rm{d}}$ for our own Galaxy (using
{\em COBE\/} data), and that of other galaxies (which use {\em IRAS\/}
measurements), are due to a cold dust component which has gone
undetected by {\em IRAS\/}.

(vii) The relationship between the mass of molecular hydrogen (as
determined from CO observations) and dust mass shows very little
intrinsic scatter ($\leq 50$ per cent) which implies that the CO to
H$_2$ conversion, `$X$', depends on metallicity in the same way as the
dust mass.

(vii) The star formation efficiency as traced by
$L_{\rm{fir}}/M_{\rm{H_2}}$ was found to be anti-correlated with
galaxy size (as measured by blue luminosity), a result also found by
Young (1999) using a different technique. The star formation
efficiency for these galaxies is higher on average than that for the
Galaxy by a factor of $\sim 3$ (13.3 for the sample compared to 4 for
the Milky Way).\\

\subsubsection*{FUTURE WORK}

Observations which would further our understanding of the properties
of dust in galaxies and refine our initial estimates of the LF are as follows:

\begin{itemize} \item Shorter wavelength (300 -- 600$\mu$m) data from
SCUBA and from SHARC on the CSO will allow us to ascertain the presence
of any cold dust component in these galaxies and whether it is
uniform (i.e. is the cold component similar in all the
galaxies or does it vary as a function of some other
property). Determining the existence of such a cold dust will allow a
much more accurate estimate of the dust mass (and hence the dust mass
function) to be made.

\item In order to resolve some of the questions unanswered by this paper we
will need to complete the survey of an optically selected sample. This
will help to constrain the behaviour of the low luminosity end of the
850$\mu$m LF (i.e. the turn-over) and also establish whether there is
a population of `cold' galaxies missing from this {\em IRAS}
sample. We will then be able to investigate the differences between
the {\em IRAS} galaxies when compared to `normal' optically selected
galaxies in terms of their dust properties.

\item The 850$\mu$m LF also requires more data points at the high
luminosity end. This sample contained only a few very luminous objects
because of the high flux limit for the BGS. Observations of a complete
sample of ULIRGS at 850$\mu$m would prove very useful in this respect.

\item More CO and HI measurements, for those galaxies in both the {\em
IRAS} and optically selected samples which do not have them already,
would increase the statistical significance of the gas and dust
comparisons. In addition, detailed mapping in both HI and CO would be
of great importance in determining the relationship of the dust to the 
different gas phases of the ISM. Work on obtaining this data is in
progress.

\item Optical imaging would enable a
comparison of the optical structures with those in the
submillimetre. Of importance is the radial profile and scale height of 
the dust compared to the stars. Extinction maps could be created from
multi-colour optical data and compared with the submillimetre maps, a
technique described by Trewhella (1998).
\end{itemize}

\section*{ACKNOWLEDGMENTS}

We would like to thank all of the support staff at the JCMT as well as
any observers who took data on our behalf when the conditions were too
poor for their own projects. In particular, we are grateful to Wayne
Holland for the 'out of hours' support after our runs were over. We
also thank Paul Alton for useful discussions and information about NGC
891. The work of L. Dunne, S. Eales, M. Edmunds, R. Ivison and
D. Clements is supported by PPARC.
